\documentclass{aa}  

\usepackage{graphicx}
\usepackage{txfonts}
\usepackage[version=3]{mhchem}
\usepackage{xcolor}

\usepackage{booktabs}
\usepackage{hyperref}

\newcommand{\fig}{Fig.}
\newcommand{\Fig}{Fig.}
\newcommand{\figref}[1]{\fig~\ref{#1}}
\newcommand{\Figref}[1]{\Fig~\ref{#1}}

\newcommand{\Tabref}[1]{Table~\ref{#1}}

\renewcommand{\eqref}[1]{Eq.~(\ref{#1})}

\begin{document} 

\title{Reaction dynamics on amorphous solid water surfaces using interatomic machine learned potentials}

\subtitle{Microscopic energy partition revealed from the \ce{P + H -> PH} reaction.}

\author{G. Molpeceres\inst{1} \and V. Zaverkin\inst{2} \and K. Furuya\inst{3} \and Y. Aikawa\inst{1} \and J. K\"astner\inst{2}}

\institute{Department of Astronomy, Graduate School of Science, The University of Tokyo, 113-0033, Tokyo, Japan\\
\email{molpeceres@astron.s.u-tokyo.ac.jp}
\and
University of Stuttgart, Faculty of Chemistry, Institute for Theoretical Chemistry \\
\email{zaverkin@theochem.uni-stuttgart.de} 
\and
National Astronomical Observatory of Japan, 181-8588, Tokyo, Japan}
        
\date{Received \today ; accepted \today}
 
\abstract
{Energy redistribution after a chemical reaction is one of the few mechanisms to explain the diffusion and desorption of molecules which require more energy than the thermal energy available in quiescent molecular clouds (10 K). This energy distribution can be important in phosphorous hydrides, elusive yet fundamental molecules for interstellar prebiotic chemistry.}
{Our goal with this study is to use state-of-the-art methods to determine the fate of the chemical energy in the simplest phosphorous hydride reaction.}
{We studied the reaction dynamics of the \ce{P + H -> PH} reaction on amorphous solid water, a reaction of astrophysical interest, using \emph{ab-initio} molecular dynamics with atomic forces evaluated by a neural network interatomic potential.}
{We found that the exact nature of the initial phosphorous binding sites is less relevant for the energy dissipation process because the nascent PH molecule rapidly migrates to sites with higher binding energy after the reaction. Non-thermal diffusion and desorption-after-reaction were observed and occurred early in the dynamics, essentially decoupled from the dissipation of the chemical reaction energy. From an extensive sampling of reactions on sites, we constrained  the average dissipated reaction energy within the simulation time (50 ps) to be between 50 and 70 \%. Most importantly, the fraction of translational energy acquired by the formed molecule was found to be mostly between 1 and 5 \%.}
{Including these values, specifically for the test cases of 2\% and 5\% of translational energy conversion, in astrochemical models, reveals very low gas-phase abundances of PH$_{x}$ molecules and reflects that considering binding energy distributions is paramount for correctly merging microscopic and macroscopic modelling of non-thermal surface astrochemical processes. Finally, we found that PD molecules dissipate more of the reaction energy. This effect can be relevant for the deuterium fractionation and preferential distillation of molecules in the interstellar medium.}


\maketitle

\section{Introduction }\label{sec:intro}

Phosphorous is an element of great relevance for the chemistry of the interstellar medium (ISM). This is due to the intrinsic connection between abiotic and prebiotic phosphorous, with abiotic molecules such as phosphine (\ce{PH3}) serving as an indicator of the potential presence of more complex oxoacids \citep{Turner2018}. Biocompatible phosphorous can be found in several essential biomolecules, such as DNA, RNA, or phospholipids. So far, phosphorous has been unambiguously detected in different regions of the ISM in the form of CP, HCP, PN, PO, or \ce{HPO} \citep{Turner1987,Guelin1990,Agundez2007, Agundez2014, Fontani2016,Lefloch2016,Rivilla2016, Rivilla2018, Ziurys2018}. However, the detection of phosphorous-bearing molecules in cold star-forming regions is limited to PN, PO, and recently, \ce{PO+} \citep{yamaguchi_detection_2011, Rivilla2016, Lefloch2016, Rivilla2018, Rivilla2020, Rivilla2022}. Despite laboratory experiments and computational simulations show that \ce{PH3} must be easily hydrogenated on dust grains \citep{Nguyen2020, Molpeceres2021b, Nguyen2021}, no single phosphorous hydride has been detected in the cold ISM \citep{Chantzos2020}. 

The sequence of hydrogenation of phosphorous atoms in the ISM is known to finalize with the formation of phosphine, \ce{PH3} \citep{Molpeceres2021b, Nguyen2021} and proceeds on interstellar dust grains through a sequence \ce{P ->[+\text{H}] PH ->[+\text{H}] PH2 ->[+\text{H}] PH3}. Competing with this sequence, there are reactions mediated by quantum tunnelling in the backward direction \ce{PH3 + H -> PH2 + H2}, \ce{PH2 + H -> PH + H2}, \ce{PH + H -> P + H2} that proceed with very low activation barriers and are very rapid, even at 10 K \citep{Molpeceres2021b}. These establish a pseudo-equilibrium that favours addition (\ce{PH3} as end molecule) in the absence of other mechanisms that reduce the intermediate species' abundance. We find chemical desorption and non-thermal diffusion among this sequence's possible mechanisms. The low binding energies (BE) of the P-bearing molecules \citep{Molpeceres2021b} combined with the confirmation of chemical desorption of phosphine under astrophysical conditions \citep{Nguyen2020, furuya_quantifying_2022} indicate that the hydrogenation sequence may be altered non-thermally such as the release of intermediates to the gas phase or by fast diffusive reactions with other atoms or radicals, promoted by the release of chemical energy. 

Regrettably, there is no possibility of determining the efficiency of non-thermal effects from static quantum chemical calculations, not only in phosphorous chemistry but also in interstellar surface chemistry. This is because the appearance of non-thermal effects requires the interplay between energy dissipation to the bulk and the conversion of the reaction energy to kinetic energy. Classical molecular dynamics simulations have been used to that end, targeting, for example, HCO formation on \ce{H2O} \citep{Pantaleone2020}, \ce{H2} formation on \ce{H2O} \citep{Pantaleone_2021}, or energy dissipation of \ce{CO2}, \ce{H2O} and \ce{CH4} on water ice after inoculation of a significant amount of energy \citep{Fredon2017, Fredon2018, Fredon2021,Upadhyay2021, Upadhyay2022}. Finally, very recently, the hydrogenation sequence of the nitrogen atom was studied on \ce{H2O} \citep{Ferrero2023}. The interaction potentials for these studies come in different flavours, such as empirical pair potentials or on-the-fly \textit{ab-initio} molecular dynamics simulations. The former's low computational cost is the latter's main drawback. Likewise, \emph{ab-initio} molecular dynamics' main advantage is that it is, in principle, general, can treat reactive systems, and new potentials do not need to be generated for each system.

In this work, we study the energy dissipation following the first hydrogenation step of the phosphorous atom (P) with a twofold aim. First of all, we study the energy dissipation dynamics of a relatively heavy radical \ce{PH} after its formation \emph{via} \ce{P + H -> PH}. The energy dissipation and reaction energy redistribution to the different degrees of freedom of the molecule is the crucial quantity behind non-thermal events and, ultimately, in the fate of phosphorous hydrides in the ISM. The second aim of this paper is to serve as a proof-of-concept for applying interatomic neural network potentials to study reactions in the context of surface astrochemistry. These potentials permit an extensive sampling of reaction trajectories while maintaining a reasonable computational cost, yet somewhat higher than empirical potentials. Using neural-network potentials opens new avenues for studying any adsorbate on any substrate, provided that the chosen reference method can reproduce the electronic structure of the system. The PH radical is especially suited as a test case for several reasons. First, its formation reaction is barrierless and exothermic ($\sim$ 305--315 kJ mol$^{-1}$). Second, the binding energy of the PH radical is relatively low, especially compared to the reaction energy, permitting the study of various outcomes after the reaction. Third, the reaction occurs on a triplet potential energy surface (PES) which facilitates the construction of the training set compared with more complex radical-radical recombinations occurring in the singlet channel. Fourth and last, PH is a sufficiently heavy molecule in which the drawbacks of using classical dynamics not accounting for nuclear quantum effects can be partially palliated.

This paper's structure is as follows. In Section \ref{sec:methods} we briefly introduce the methodology employed in the paper and the particularities of the training set used to train the interatomic neural network potential. In Section \ref{sec:results}, we present the results derived from the present study, starting with the adsorption energetics of the P atom at our level of theory and continuing with exploratory molecular dynamics simulations of different binding sites. Later, we focus on extensive sampling, extracting statistical values on energy dissipation and translational kinetic energy of the newly formed \ce{PH} radical. Finally, we discuss our results under an astrochemical prism in Section \ref{sec:Discussion}, and include our results in an astrochemical model of a molecular cloud.

\section{Methods} \label{sec:methods}

\subsection{Interatomic Potential Model}

In this work, we have developed an entirely reactive machine-leaned interatomic potential (MLIP) to study the reaction dynamics for forming the PH molecule on top of amorphous solid water (ASW). Our approach resembles Born--Oppenheimer molecular dynamics at a fraction of its cost. In our simulations, we must sample many trajectories with different reaction outcomes for sufficiently long time scales. Moreover, we need sufficient energy conservation to obtain correct dynamics, especially during the initial step of the reaction. Small integration time steps ensure the desired energy conservation, \emph{vide infra}. We have achieved all these prerequisites by combining ample sampling with a Gaussian-Moment Neural Network (GMNN) potential \citep{Zaverkin2020, Zaverkin2021} trained on energies and atomic forces obtained from density functional theory (DFT). The GMNN source code is available free-of-charge at \href{https://gitlab.com/zaverkin\_v/gmnn}{gitlab.com/zaverkin\_v/gmnn}.

To calculate reference energy and atomic force values, we use the DFT method PBE-D3BJ/def2-TZVP \citep{Perdew1996, Weigend2005, Grimme2011} as it provides similar energetics to those obtained by \cite{Molpeceres2021b, Nguyen2021}. The \ce{P + H -> PH} reaction energy (without zero-point vibrational energy, ZPVE) in the gas phase is within 9 kJ mol$^{-1}$ of reference UCCSD(T)-F12/cc-pVTZ-F12 calculations, $-$314 \emph{vs.} $-$305 kJ mol$^{-1}$. We studied the reaction in the triplet channel since the quintet channel was found to be repulsive in an earlier publication \citep{Molpeceres2021b}. The training data set generated to train the GMNN model comprises 11,249 structures. The training set size required to achieve the accuracy of 1 kcal\,mol$^{-1}$ with respect to the reference electronic structure method is smaller than in our previous works \citep{Molpeceres2020_2, Molpeceres2021a, Zaverkin2021b} owing to the recent improvements in our model \citep{Zaverkin2021}. The electronic structure calculations were carried out using the ORCA (v 5.0.3) code \citep{Neese2012, Neese2020}.

A summary of the structures which comprise the training set can be found in \Tabref{tab:training_set}. Like in our previous work \citep{Molpeceres2020_2, Molpeceres2021a}, we generated the training structures by running \textit{ab-initio} molecular dynamics at exploratory, computationally cheaper levels of theory. Here, we used a combination of HF-3c \citep{Sure2013} for the exploration of reactions on clusters and GFN2-xTB \citep{Bannwarth2019}, as well as  GFNFF \citep{Spicher2020} for studying water clusters and isolated adsorbates on water clusters. The training set includes sub-sets tailored to reproduce the water-water interaction, for which molecular dynamics in the canonical (NVT) ensemble was carried out, and other sub-sets simulating collision and reactions, where a microcanonical ensemble (NVE) was used.

\begin{table*}[t]
\caption{Composition of the data set used for training the machine-learned interatomic potentials. Note that the propagation method is only used for sampling geometries. For these structures, energies and forces are later calculated at the PBE-D3BJ/def2-TZVP level of theory.}  
\label{tab:training_set} 
\centering
\resizebox{\linewidth}{!}{\begin{tabular}{lllll}
\toprule
Label & Number of Points & Ensemble & Temperatures (K) & Propagation Method \\
\midrule
31 \ce{H2O + H}                 & 240   & NVT & 300/150/50                  & GFN2-xTB      \\
31 \ce{H2O + P}                 & 240   & NVT & 300/150/50                  & GFN2-xTB      \\
31 \ce{H2O + PH}                & 240   & NVT & 300/150/50                  & GFN2-xTB      \\
46 \ce{H2O + H}                 & 600   & NVT & 300/150/50                  & GFN2-xTB      \\
46 \ce{H2O + P}                 & 600   & NVT & 300/150/50                  & GFN2-xTB      \\
46 \ce{H2O + PH}                & 600   & NVT & 300/150/50                  & GFN2-xTB      \\
74 \ce{H2O}                     & 800   & NVT & 100/500$^1$                 & GFN-FF        \\
\ce{12H2O + H} Collision        & 500   & NVE & 100 (E$_{\ce{H}}$=1.5 eV)   & GFN2-xTB      \\
\ce{12H2O + P} Collision        & 500   & NVE & 100 (E$_{\ce{P}}$=0.025 eV) & GFN2-xTB      \\
Long-range \ce{P}$^2$           & 400   & N/A & N/A                         & N/A           \\
\ce{20H2O + P + H} Reaction$^3$ & 6927  & NVE & 30                          & HF-3c         \\
\midrule    
Total & \multicolumn{4}{l}{11,647 (11,249)$^4$}\\
\bottomrule
\multicolumn{5}{l}{$^1$\footnotesize{A spherical wall potential is applied to ensure the structural integrity of the cluster.}} \\
\multicolumn{5}{l}{$^2$\footnotesize{Interaction of P with the cluster at the distance of the cutoff radius (5.5~\AA{}, required for binding energies).}} \\
\multicolumn{5}{l}{$^3$\footnotesize{The reaction part comprises reactions in weak, medium and high binding sites at different P-H internuclear distances (3.0--4.5 \AA{}).}} \\
\multicolumn{5}{l}{$^4$\footnotesize{398 structures have been removed from the training set (some distances between neighboring atoms are larger than the selected cutoff-radius).}}
\end{tabular}}
\end{table*}

Three independent MLIP models were then trained for 1000 epochs. All models have been trained on 9000 structures from the data set in \Tabref{tab:training_set}, while other 1500 structures have been reserved for the early stopping technique \citep{Prechelt2012}. The achieved mean absolute error with respect to the reference DFT level is about 0.5 kcal$\,$mol$^{-1}$ and 0.3 kcal$\,$mol$^{-1}$$\,$\AA$^{-1}$ for total energies and atomic forces, respectively. The provided errors have been evaluated on the test data set consisting of the remaining 749 structures. We then used an ensemble of the three models to predict energies and atomic forces while running our simulations. The cutoff radius is set to 5.5~\AA{} because we are interested in short and medium-range interactions of the adsorbates with the surface. All other relevant hyper-parameters used for training our NN potentials can be found in the original work \citep{Zaverkin2021}.

The advantage of using an ensemble is two-fold. First, lower error values in the potential energy can be achieved, and the predicted atomic forces are much smoother than for a single model. The latter is crucial for energy conservation. Second, the variance or disagreement between the three independent models estimates the uncertainty in predicted energy and atomic force values. Thus, using an ensemble of at least three NN potentials provides the means of running a more stable molecular dynamics simulation and allows for estimating uncertainties in predicted values such as binding energy, reaction energy and reaction profiles. Finally, a detailed analysis of the accuracy of the employed NN ensemble is provided in Appendix \ref{sec:appendix}.

\subsection{Water ice surface model} 

With the MLIP trained, we created a slab model for an amorphous solid water surface by running heating and cooling cycles of molecular dynamics simulations. Very briefly, from a randomly, pre-optimized and fully periodic packed simulation cell with 500 water molecules at $\rho$=0.998 g cm$^{-3}$ we run NVT dynamics at 300~K for 100~ps using a Langevin thermostat to control the temperature (friction coefficients of 0.02 ps$^{-1}$ and 0.5 fs timestep). Five different snapshots along the trajectory are chosen and later quenched to 10~K for 10.0 ps (friction coefficients of 0.05 ps$^{-1}$ and 0.5 fs timestep), serving as initial models for the subsequent reactivity studies. This system is periodic in the $X$ and $Y$ directions but not in the $Z$-direction. We propagate our dynamics in periodic systems while having trained on cluster data. We refer to our recent publication for details on this procedure \citep{Zaverkin2022}.

We fixed the bottom layers of the simulation cell in the $Z$ direction to simulate a bulk structure.
Once we had the structural models for the ice ready, we sampled the binding energy distribution of the P atom by placing P atoms atop the surface and relaxing the structure. Finally, from representative configurations of the binding energy distribution (see Section \ref{sec:binding}), we start NVE simulations to trace the reaction dynamics. The chemical reaction under investigation is \ce{P + H -> PH}. The incoming H atoms are placed around the pre-adsorbed P atom in different binding sites at a distance of 3.5~\AA{}. For all simulations, we run NVE dynamics for 50~ps using an integration timestep of 0.25 fs. All molecular dynamics calculations were performed with the ASE simulation package \citep{HjorthLarsen2017}.

\section{Results} \label{sec:results}

\subsection{Binding energy distributions calculation} \label{sec:binding}

\begin{figure}
        \centering
        \includegraphics[width=8cm]{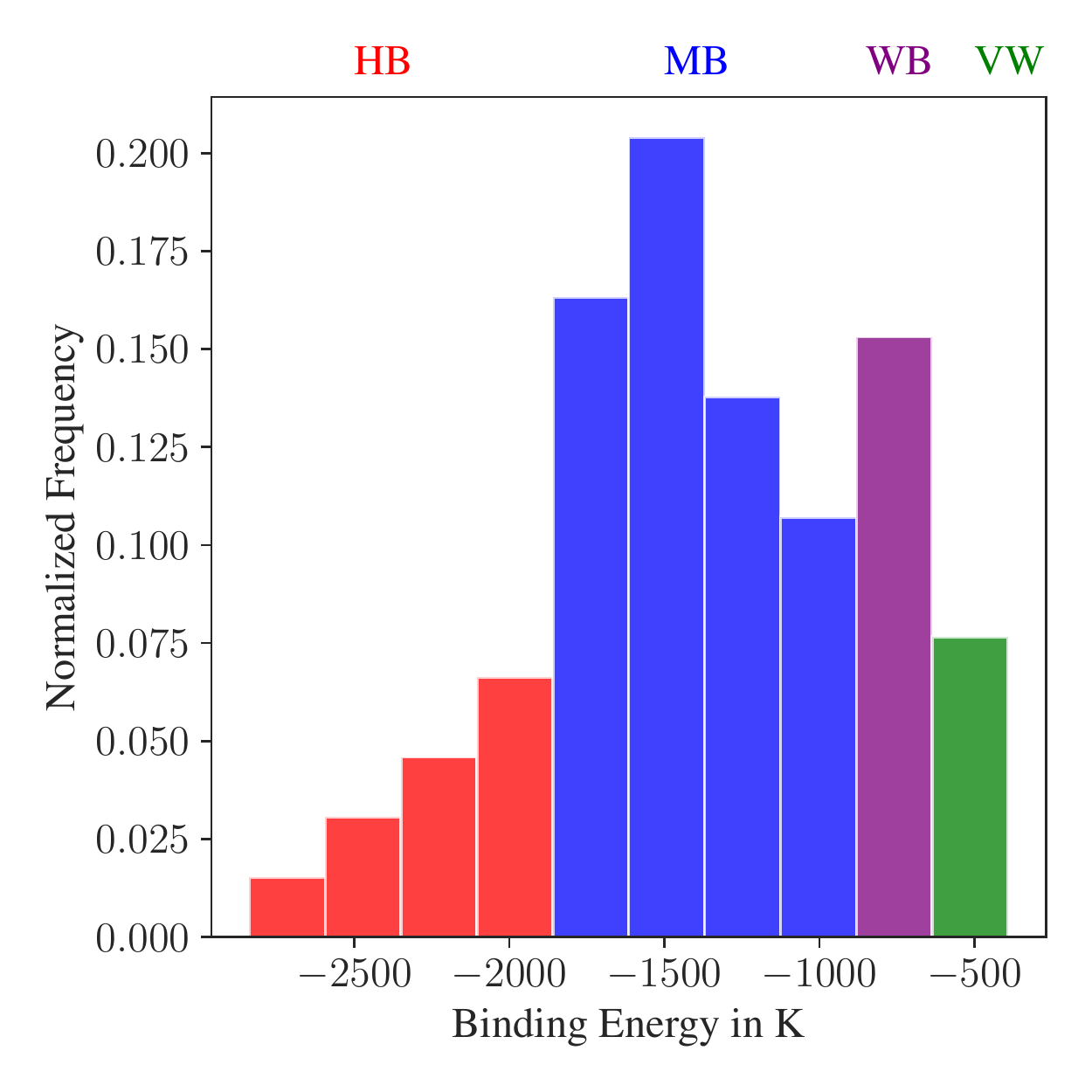}
        \includegraphics[width=8cm]{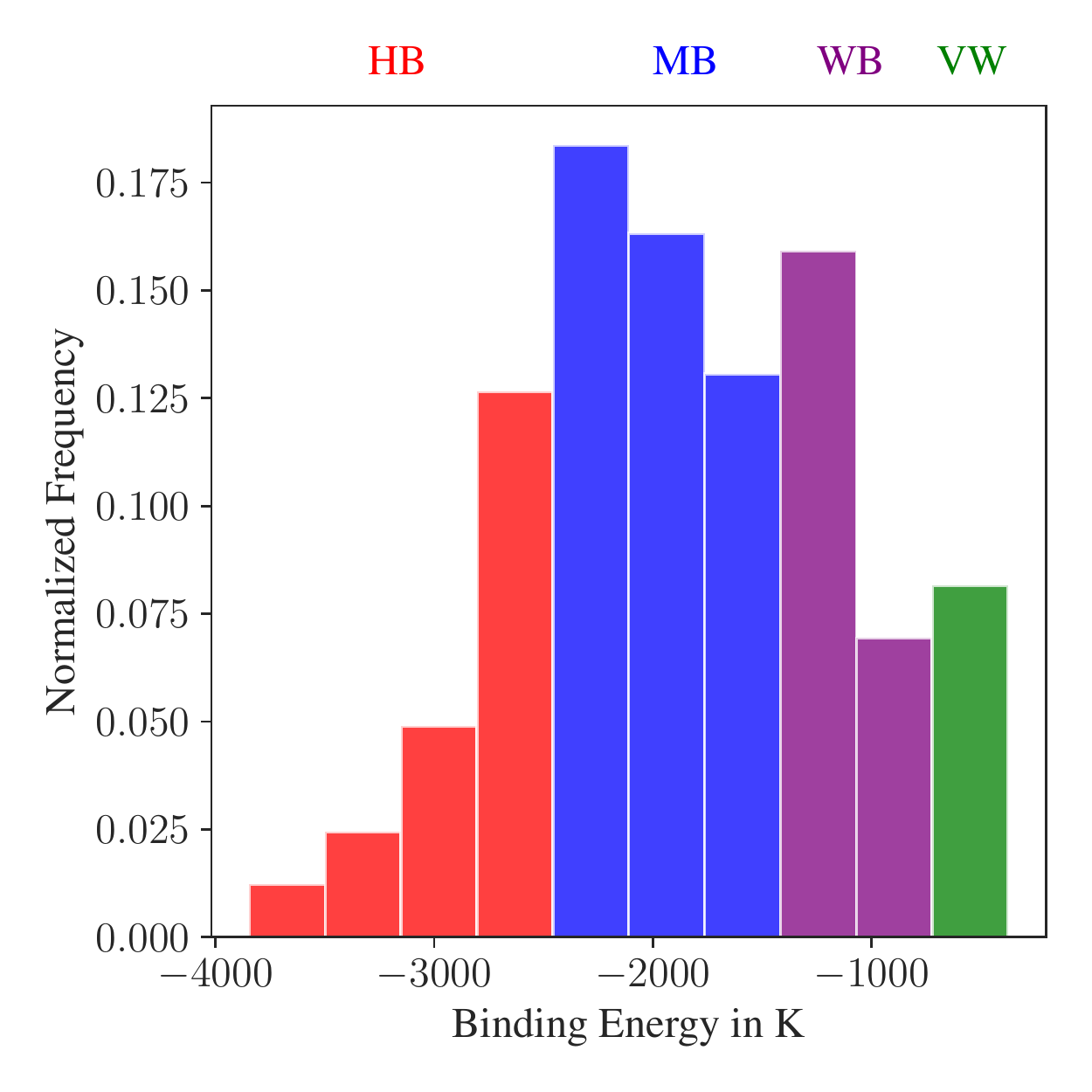}
        \caption{Distribution of the binding energies of P (Top) and PH (Bottom) on ASW. Each bar is coloured according to the approximated attributed binding site. The colour code is high binding (HB) red, medium binding (MB) blue, weak binding (WB) purple, very weak binding (VW) green. We sampled 245 binding sites to obtain these distributions.}
    \label{fig:binding}
\end{figure}

The binding energy distribution of the P atom and the PH molecule on ASW, computed with the MLIP, is very similar to our previous work \citep{Molpeceres2021b} using more sophisticated density functionals explicit calculations. To ease the discussion in \figref{fig:binding}, we identify several segments. One corresponds to high binding (HB), at values of $>$ 120\% of the average binding energy (which for P is at 1371~K, equivalent to 1241~K of our previous work \citep{Molpeceres2021b} and other works \citep{Nguyen2021}). Medium binding (MB) sites, around the average binding energy, weak binding (WB) sites at 70--40\% of the average binding energy, and very weak binding (VW) sites at $<$ 40\% of the average binding energy. Overall, the binding energy of P on ASW is relatively small for an atom as heavy as phosphorus. The average binding energy for PH is 1843 K, also in good agreement with our previous work (1616 K; \cite{Molpeceres2021b}), especially considering that we omitted ZPVE in the distributions because we cannot capture the ZPVE during the reaction dynamics.

\begin{figure}
    \centering
    \includegraphics[width=4cm]{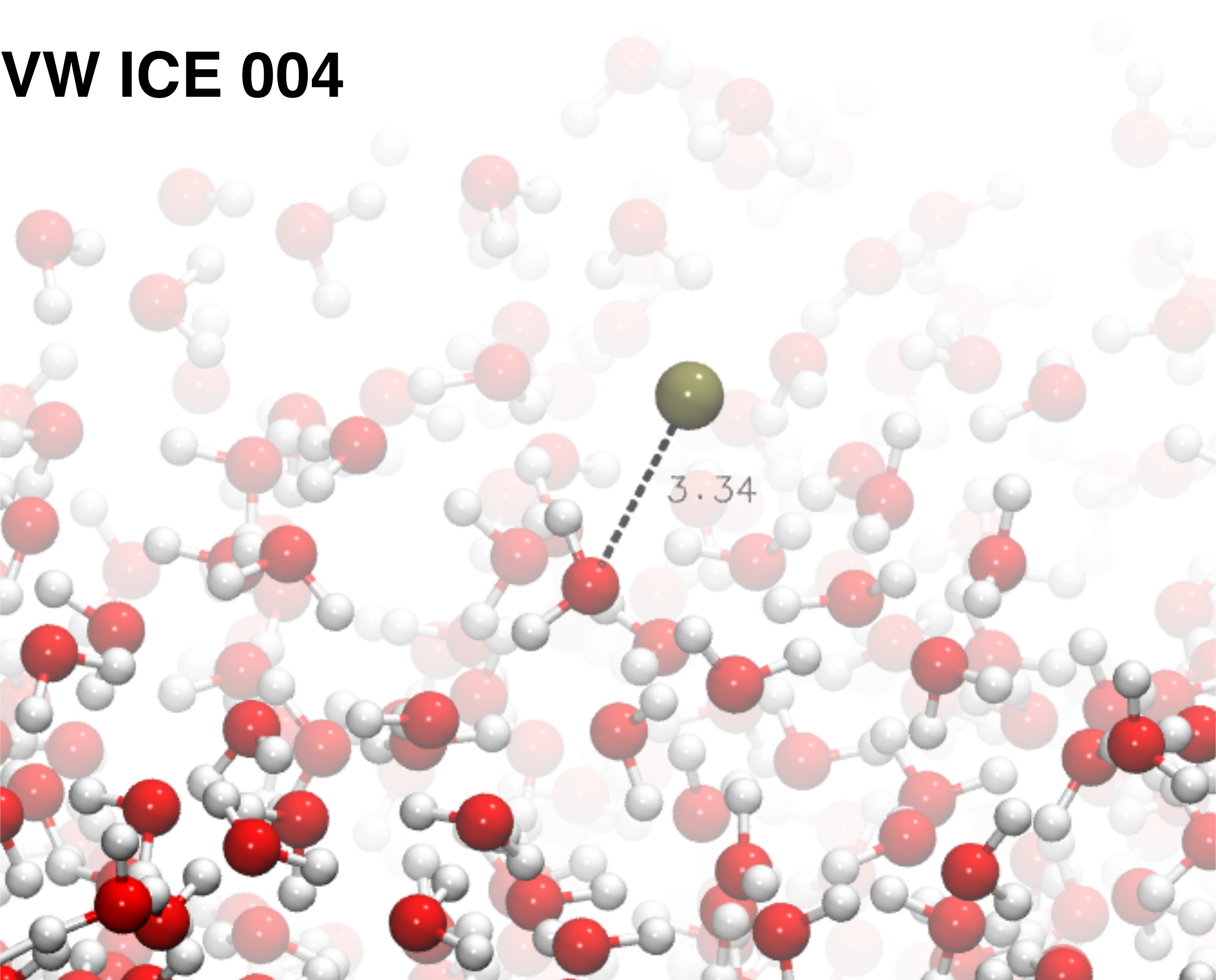}
    \includegraphics[width=4cm]{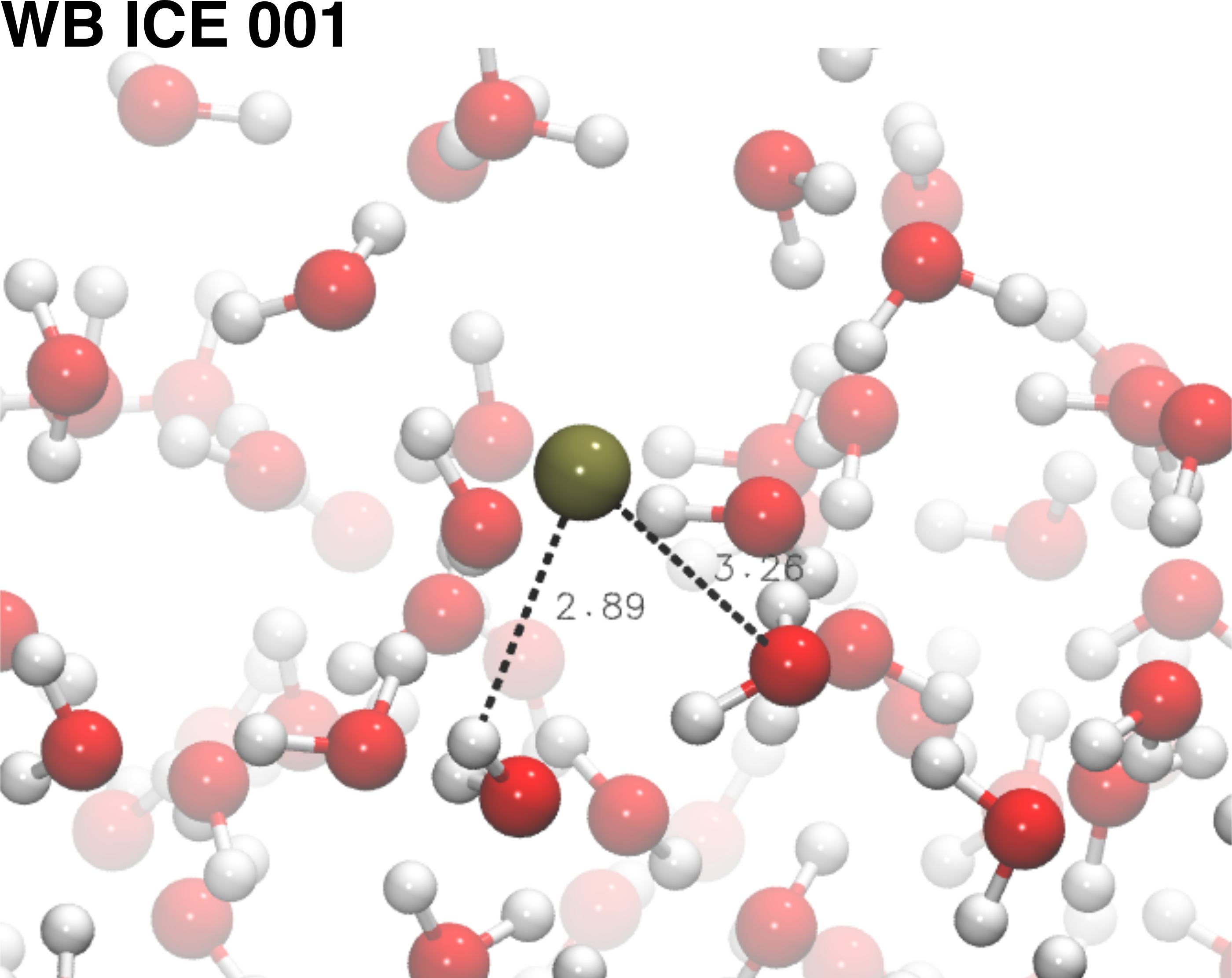}
    \vfill
    \includegraphics[width=4cm]{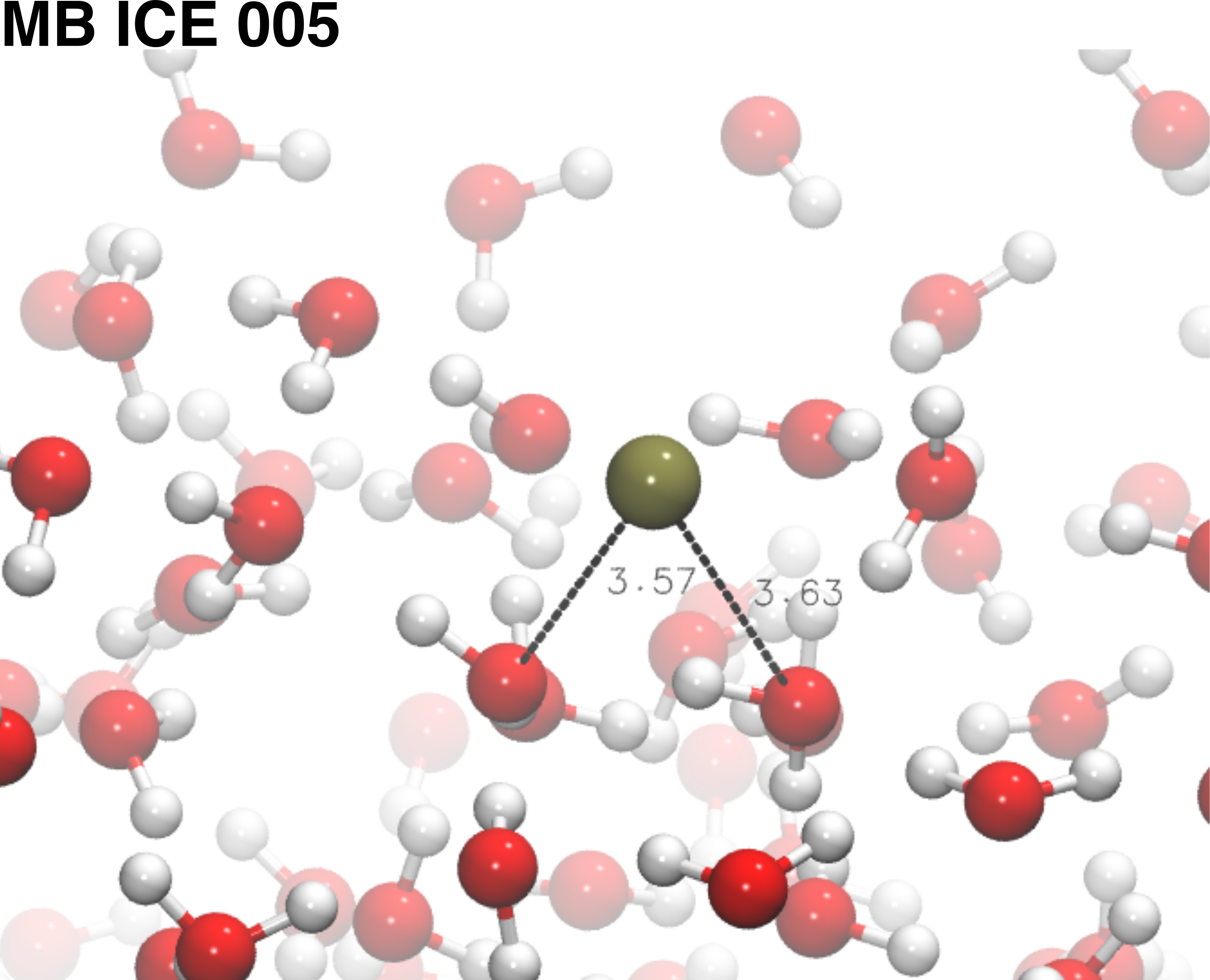}
    \includegraphics[width=4cm]{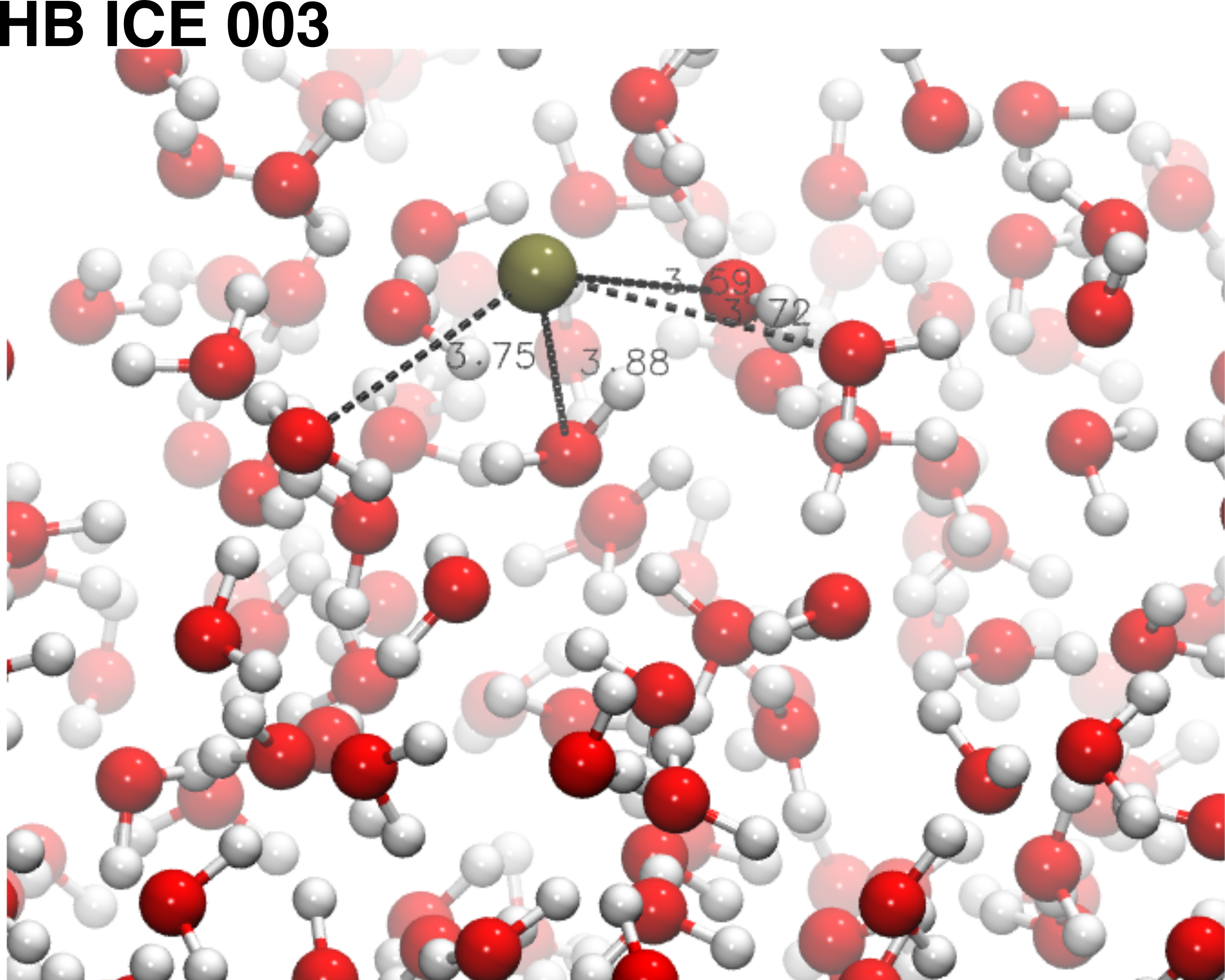}
    \caption{Local environments of the P atom prior to the reaction with H. The models portrayed in this figure come from different ice models.}
    \label{fig:pfigures}
\end{figure}

 As mentioned above, we selected four initial configurations to study the reaction from the distribution of binding energies for the P atom. The local environments for each configuration in one of our ice models are depicted in \figref{fig:pfigures}. The figure illustrates that more oxygen neighbours increase the binding energy, as predicted previously \citep{cuppen_kinetic_2013}. On the contrary, interactions with H atoms lead to weaker binding. The first part of our study evaluated the binding site's dependence on the dynamics, which was sampled for 19 trajectories because we faced difficulties finding a VW situation for one of the ice models. As mentioned, the P--H$_\mathrm{incoming}$ distance was set to 3.5~\AA{}. We run a short ($\sim$ 2~ps) NVT simulation to initialize the velocities of the water molecules to 10 K before the production run. During this short pre-equilibration, both P and H positions can relax, but we enforce a constraint in the internuclear distance to the above-mentioned value.

\begin{figure}
    \centering
    \includegraphics[width=8cm]{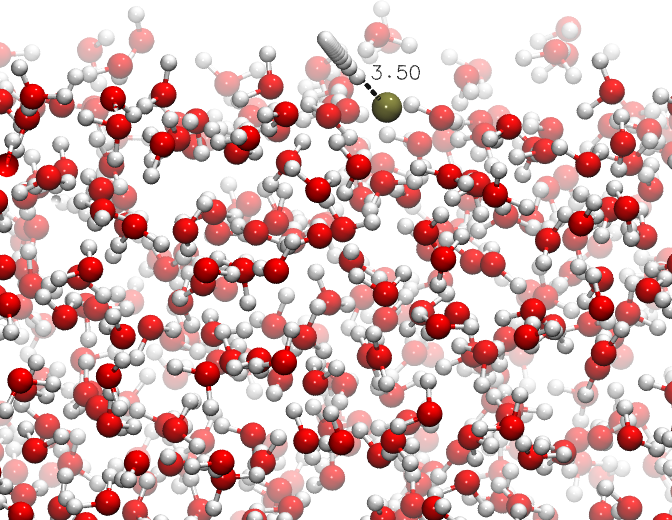}
    \caption{Onset for the title reaction in our simulations. The figure represents around 50 fs of the total dynamics (50 ps). The distance of 3.5 \AA{} represents the P--H initial distance.}
    \label{fig:onset}
\end{figure}

\subsection{Exploratory PH reaction dynamics} \label{sec:exploration}

\begin{figure*}[ht]
    \centering
    \includegraphics[width=8cm]{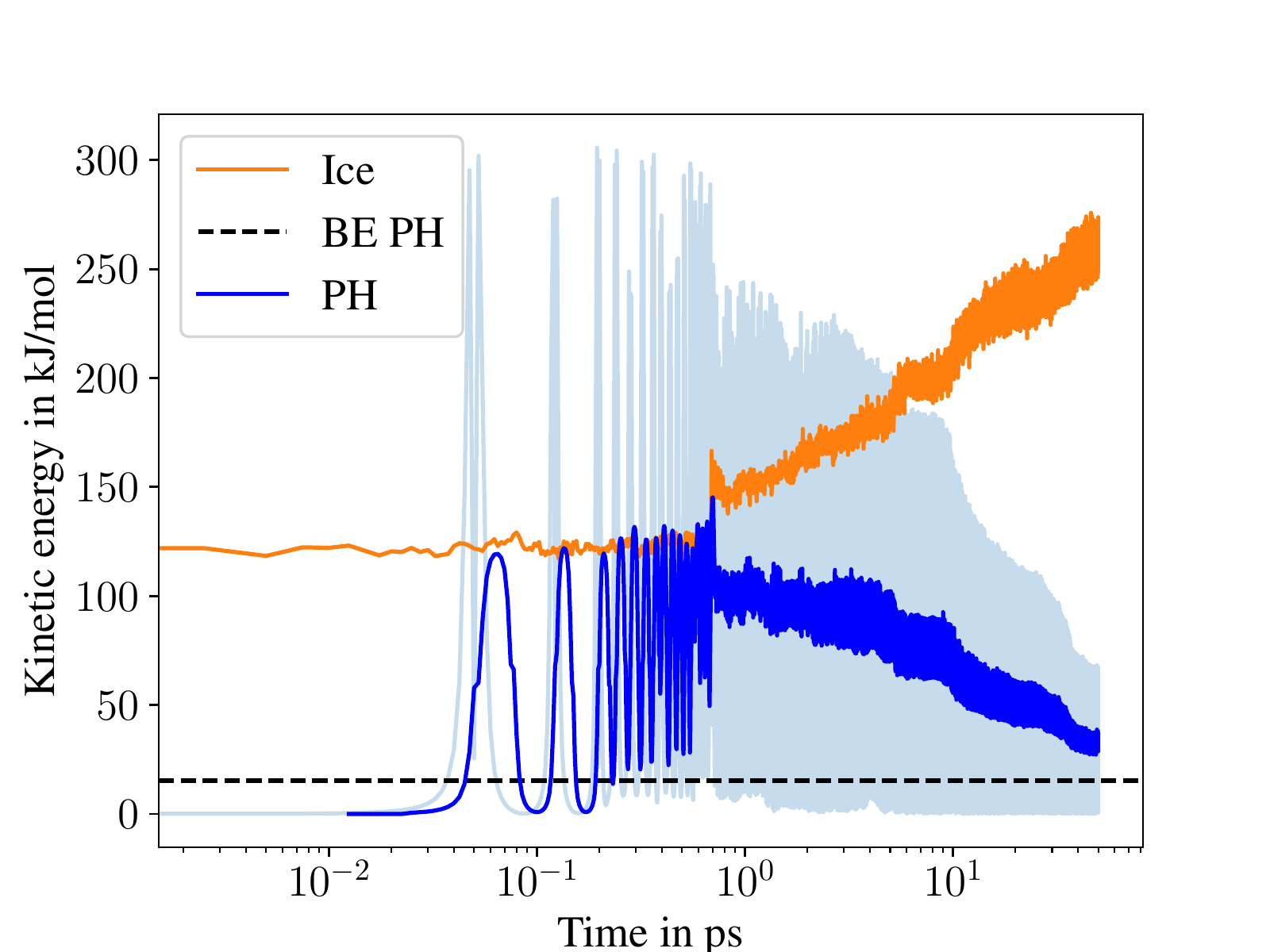} 
    \includegraphics[width=8cm]{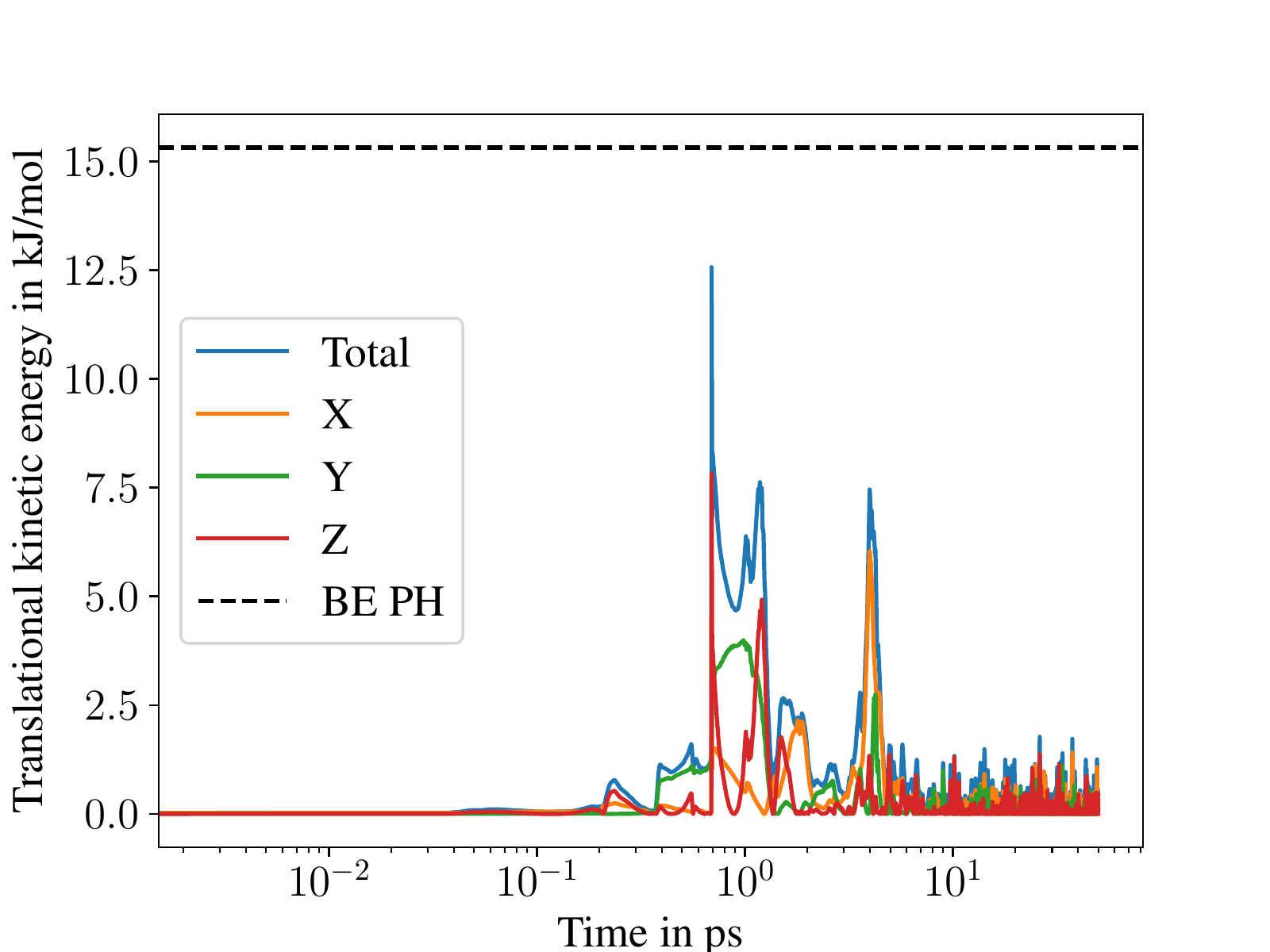}
    \vfill
    \includegraphics[width=8cm]{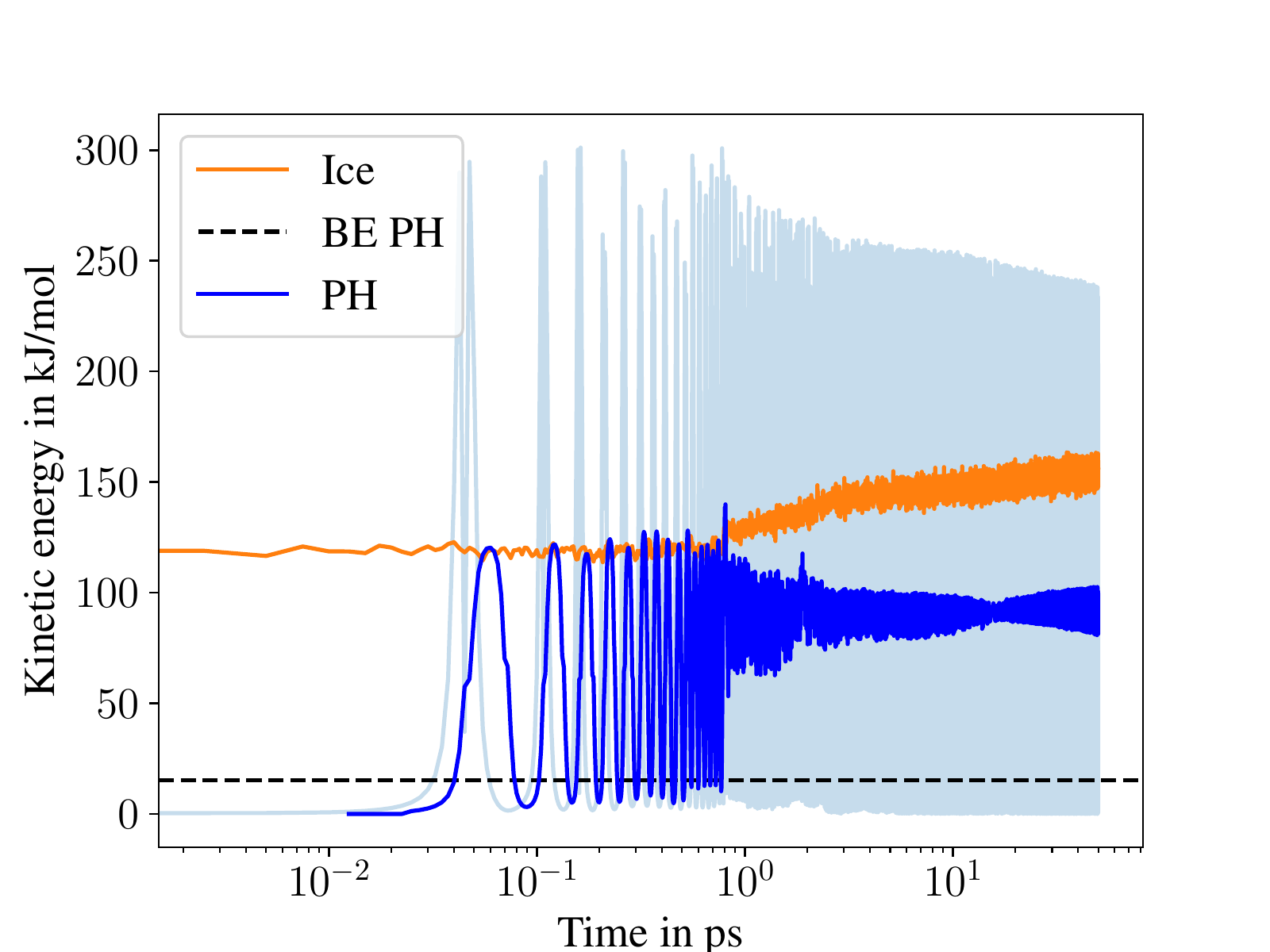} 
    \includegraphics[width=8cm]{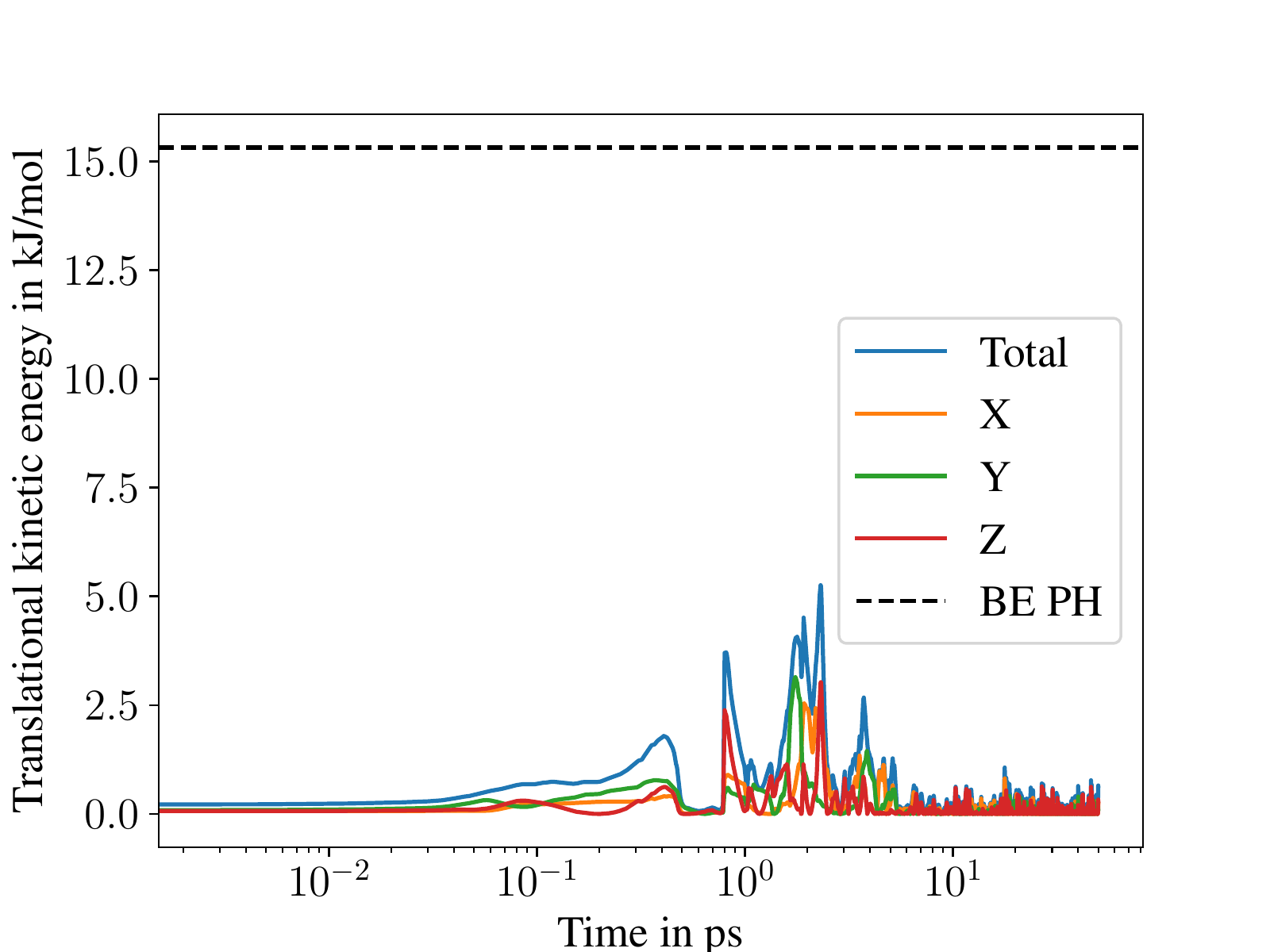}
    \caption{Evolution of the total kinetic energy of water ice and PH molecules (left panels) and PH translational energy (right panels) for two \ce{P + H -> PH} reaction trajectories, one with significant energy dissipation in the timelapse of the simulation (BE=739 K, top panel) and without significant energy dissipation (BE=805 K, bottom panel). The pale blue colour in the left panels shows the instantaneous kinetic energy of the PH molecule. The dark blue colour shows the running average evaluated over a window size of 25 fs. The black dashed line in both panels depicts the average PH binding energy. $X$, $Y$, and $Z$ indicate the component along the translational energy is distributed, with $Z$ being the normal to the surface. }
    \label{fig:kinetic_ice}
\end{figure*}

An example of the onset of the NVE dynamics for a binding site is presented in \figref{fig:onset}. From each of the trajectories, we monitored the evolution of the kinetic energy of the nascent PH molecule along the trajectory and the kinetic energy of all the water molecules within the ice. Two exemplary kinetic energy profiles for the reaction can be found in \figref{fig:kinetic_ice}, left panels. For comparison, we have selected two trajectories with very similar binding energies but different energy dissipation behaviour. In the case shown on top, the ice structure accommodates the excess reaction energy into the ice lattice quite fast, whereas, in the one shown at the bottom, most of the reaction energy remains longer as vibrational and rotational energy within the molecule. While this does not mean that energy dissipation does not occur, it hints at different dissipation timescales but without a significant dependence on the binding site.

To check the dependence of the energy dissipation process on the initial binding site, we have gathered the outcome of all these trajectories and compared their initial P binding energy with their final PH binding energy. The results can be found in \Tabref{tab:initialBEfinalBE}. We represent the population of P atoms in a given binding site before and after the reaction, along with the fraction of dissipated energy for each binding site in each ice model. The fraction of dissipated energy is calculated as:

\begin{table}[t]
\caption{Initial and final binding sites of our exploratory simulations for the \ce{P + H -> PH} reaction on ASW, including all types of binding sites along with the fraction of the reaction energy dissipated into the ice bulk for each trajectory. }  
\label{tab:initialBEfinalBE} 
\centering  
\resizebox{\linewidth}{!}{\begin{tabular}{cccccccc}    
\toprule
& \multicolumn{2}{c}{Pop. Number} &
\multicolumn{5}{c}{$\mathcal{F}$ for Ice Model} \\
\midrule
Site & P (Initial) & PH (Final)$^1$ & 001 & 002 & 003 & 004 & 005 \\
\midrule
VW & 4 & 0 & N/A$^2$ & 0.55 & 0.93 & 0.27$^3$ & 0.52 \\
WB & 5 & 0 & 0.78  & 0.64 & 0.35 & 0.43 & 0.33 \\
MB & 5 & 6 & 0.85 & 0.48 & 0.51 & 0.51 & 0.74 \\
HB & 5 & 12 & 0.49 & 0.65 & 0.53 & 0.66 & 0.27 \\
\bottomrule  
\multicolumn{8}{l}{$^1$\footnotesize{One trajectory is desorbing. Not included in the counting.}} \\
\multicolumn{8}{l}{$^2$\footnotesize{No VW binding site was found in our sampling.}} \\
\multicolumn{8}{l}{$^3$\footnotesize{Desorbing trajectory.}} \\
\end{tabular}}
\end{table}

\begin{equation}
    \mathcal{F} = \dfrac{\left| T_\mathrm{PH}^\mathrm{final} - T_\mathrm{PH}^\mathrm{max} \right|}{T_\mathrm{PH}^\mathrm{max}},
\end{equation}
where $T_\mathrm{PH}^\mathrm{final}$ is the final kinetic energy averaged over the last 2.5 ps (to avoid oscillations) of the PH molecule and $T_\mathrm{PH}^\mathrm{max}$ is the maximum of the kinetic energy of PH along the trajectory. Both quantities are calculated from the running averages with a window size of 25 fs. 
We found different values of $\mathcal{F}$ for different binding sites and ice models. However, the dependence on the different binding sites is unclear, indicating that further sampling is required. We increased the sampling for MB and HB sites in Section \ref{sec:MandH}. 

The lack of correlation between $\mathcal{F}$ and the binding site can be found by looking at the different populations of P binding sites before the reaction and PH binding sites after the reaction shown by the second and the third column in \Tabref{tab:initialBEfinalBE}. While we start our sequential simulations with P atoms in VW and WB binding sites along with MB and HB, we find that PH populates only HB and MB sites at the end. This is an essential conclusion of our work, pointing to the fact that nascent molecules arising from exothermic radical-radical recombination are unlikely to be found in the weak binding sites on a pristine surface and affecting the (effective) distributions of binding energies recently reported in the literature for low-temperature studies \citep{Ferrero2020, Duflot2021, Molpeceres2021b, bovolenta2022}. The population of HB and MB sites comes as a result of non-thermal diffusion of the nascent PH molecule that uses part of the reaction energy to freely roam the surface until landing on an HB or MB site.

We also evaluated the translational energy (part of the kinetic energy corresponding to the translational motion of the whole PH molecule) depicted in the right panels of \figref{fig:kinetic_ice} as the instantaneous kinetic energy (no running average). It is calculated as the kinetic energy of the centre of mass (T$_{\text{COM}}$). From the right panels of \figref{fig:kinetic_ice}, we observe that, despite having similar P binding energies, these PH molecules acquire different translational energies. Both roam the surface before thermally equilibrating and land at a site where they remain for the rest of the simulation. The maximum of the translational energy is, on average, only between 1 and 5\,\% of the reaction energy (-314 kJ mol$-{1}$, as a univocal value), with significant variability. Later we incorporate these values into astrochemical models. 

The key to energy dissipation is the binding site where the particles land rather than the binding site where the molecule is formed, in contrast to non-thermal events (like non-thermal diffusion) that depend on the initial site. That conclusion highlights the importance of the ice morphology in the whole energy dissipation \emph{vs.} non-thermal processes interplay. Non-thermal events are characteristic of VW, and WB sites, where P (and PH) interact with fewer neighbours. Hence, a molecule formed in a region with a low density of water molecules will be more likely to experience non-thermal diffusion or desorption. It is essential to mention that even starting the reaction from HB and MB sites, the newly formed PH molecule still may non-thermally diffuse to other (different) HB and MB sites. The fraction of PH molecules diffusing from MB and HB sites can be as low as 7\% for HB, indicating the correlation between the binding energy and non-thermal events.

\begin{figure}[ht]
    \centering
    \includegraphics[width=8cm]{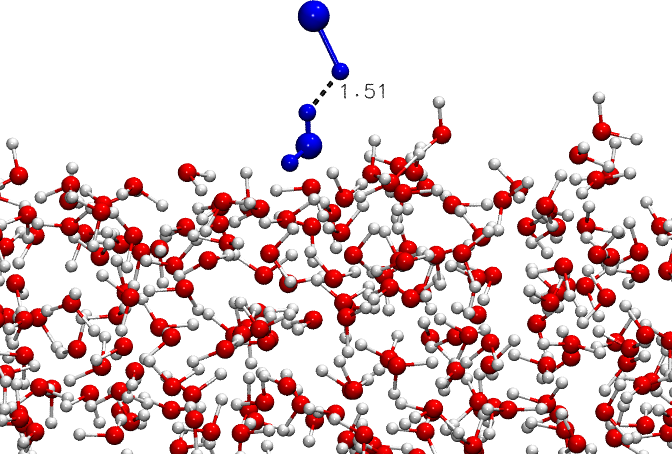}
    \vfill
    \includegraphics[width=8cm]{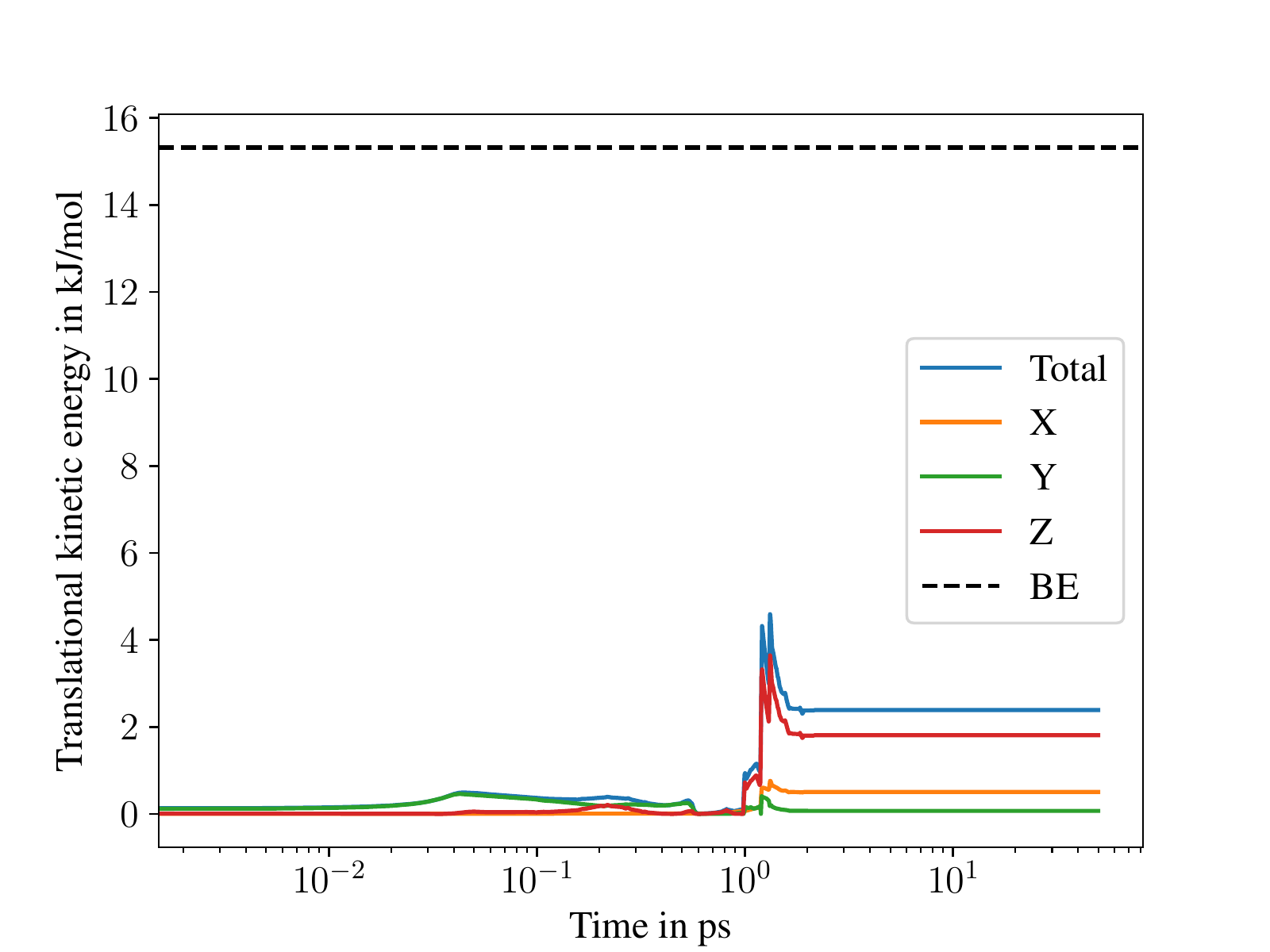} 
    \caption{Top panel: Molecular dynamics frame prior to chemical desorption. Note that even in such repulsive parts of the PES, our potential only has an uncertainty of around 10 \% between models, indicating high-quality predictions of the PES also during/throughout the reaction. Bottom panel: Translational kinetic energy of the respective trajectory.}
    \label{fig:desorption_ice}
\end{figure}

In one out of the 19 trajectories, we observed chemical desorption. We can attribute the mechanism of desorption to the interaction of a highly rotationally excited PH molecule with a dangling O--H bond during an instantaneous exploration of a very repulsive part of the potential (short H--H distance of 1.51~\AA), counteracted by momentum conservation and desorption, see \figref{fig:desorption_ice}. Another important observation of the present work is that the type of non-thermal process, either diffusion or desorption, is determined early in the dynamics (1--10 ps, inferred from the time at which peaks in the translational energy of \figref{fig:kinetic_ice} and \figref{fig:desorption_ice} take place) indicating essentially different timescales between non-thermal processes and energy dissipation. The latter, for example, can take up to a few nanoseconds,e.g., for \ce{CO2} molecules \citep{Upadhyay2022}. In \cite{Upadhyay2022} the authors also found that there are two components for the energy dissipation, one rapid, occurring in tens of picoseconds, which is coherent with our observed behaviour in Table \ref{tab:initialBEfinalBE} for $\mathcal{F}$. The second one, occurring in the nanosecond scale is not captured in our simulations because of the simulated time-window. We note that the PH molecule of the desorbing trajectory possesses a translational kinetic energy three times lower than the average binding energy for the PH radical on ASW. This shows that the likelihood of chemical desorption, like adsorption or diffusion, depends on the whole distribution of binding energies, rather than only on its average.

Our results are also in line with recent results on the \ce{N + 3H -> NH3} system, with dissipation fractions between 58 and 90\% within the first picoseconds of the reaction dynamics \citep{Ferrero2023}. We observe a similar qualitative behaviour at longer time scales (e.g. Figure \ref{fig:kinetic_ice}, left panels and Table \ref{tab:initialBEfinalBE}), which can be attributed to the different masses and vibrational frequencies of PH and NH.

\subsection{Extended sampling} \label{sec:MandH}

In the previous section, we determined that the timescales for non-thermal diffusion and desorption are different compared to the energy dissipation. Based on that, we also conclude that, except for the trajectories leading to immediate desorption of PH, the MB and HB are especially interesting for investigating/evaluating energy dissipation because, after the few picoseconds of a simulation, they are the only sites with a PH population different from zero (\Tabref{tab:initialBEfinalBE}), whereas WB sites are the most interesting from a chemical desorption perspective. P is a physisorbed species with relatively low binding energy, so VW binding sites are most likely transient.

To extract reliable statistics on $\mathcal{F}$ and related quantities, namely the maximum translational energy 
($T_{\text{COM}}$),
the linear displacement ($d$) of the PH molecule after formation, and the number of trajectories with significant non-thermal diffusion (N$_{\text{diff}}$, defined as the fraction of trajectories with $d> 2$~\AA), we have carried out an extensive sampling of MD trajectories (358 trajectories, similar to the ones presented before) starting only from MB, HB, WB situations. We observed explicit chemical desorption in three of 358 trajectories, all of which have been initialized from a WB site. These results correspond to about 1\% and the 3~\% of total chemical desorption probability and chemical desorption probability coming from WB sites, respectively. The latter number is similar to the one found for \ce{PH3} (a similar system with more degrees of freedom) in experiments \citep{Nguyen2020}, suggesting that the experimental observations may also come from WB sites. Note, however, that the sampling is not enough to guarantee a statistical match between theory and experiments, summed to the inherent differences between the PH and \ce{PH3} system. That is why we refer to our discussion in terms of translational energy in Section \ref{sec:Discussion}.

The significant quantities for these trajectories are shown in \Tabref{tab:bigSamplingH}. There, we also include statistics of further 233 trajectories, in which the reactive H atom was replaced by D for MB and HB sites (i.e. MB-D and HB-D). We did not include the study with deuterium in WB sites because of the computational expense of the respective calculations. With the deuterium substitution, we want to observe whether the different vibrational properties of \ce{PH} and \ce{PD} affect the dynamics of the nascent molecule. For example, \ce{PD} has its fundamental vibrational mode in the range of water bendings ($\sim$ 1600 cm$^{-1}$), unlike PH that appears in a clean region of the spectra at 2218 cm$^{-1}$, which can facilitate the energy dissipation through the bending modes of water. 

\begin{table}[t]
\caption{Number of trajectories, average dissipation fraction ($\overline{\mathcal{F}}$), average translational kinetic energy ($\overline{T}_{\text{COM}}$, in kJ mol$^{-1}$), the fraction of trajectories exhibiting non-thermal diffusion ($N_{\text{diff}}$, see text) and average diffusion distance for the diffusing trajectories ($\overline{d}$, in \AA, the standard deviation in parentheses).}  
\label{tab:bigSamplingH} 
\centering  
\resizebox{\linewidth}{!}{\begin{tabular}{cccccc}
\toprule
Site & $N_\mathrm{traj}$ & $\overline{\mathcal{F}}$ & $\overline{T}_{\text{COM}}$ & N$_{\text{diff}}$ & $\overline{d}$ \\
\midrule
MB & 125 & 0.67 & 5.8 & 0.81 & 5.6(2.5)    \\
HB & 108 & 0.55 & 3.3 & 0.07 & 5.0(2.8)     \\
MB-D & 125 & 0.73 & 5.3 & 0.64 & 5.1(2.6) \\
HB-D & 108 & 0.61 & 2.7  & 0.10 & 4.4(2.5) \\
WB$^1$   & 122  &  0.62  &  9.1 & 1.00   &  6.9(3.6)        \\
\bottomrule  
\multicolumn{6}{l}{$^1$\footnotesize{Three trajectories leading to chemical desorption are excluded.}}
\end{tabular}}
\end{table}

The analysis of the trajectories for the MD runs reveals several important conclusions. First, we have found that the type of binding site significantly affects the energy dissipation in the considered timescales. This fact could be inferred from previous studies \citep{Garrod2007,Minissale2016b, Fredon2018, Fredon2021} where the efficiency of chemical desorption is a function of BE. However, as previously indicated for thermal desorption studies \citep{Molpeceres2021b, Ferrero2022, Molpeceres2022b}, we must emphasize that considering the whole distribution of binding sites is essential to unravel the role of non-thermal mechanisms. Here we found that the binding energy anticorrelates with the fraction of energy dissipated into the lattice and the acquired translational energy of the nascent molecule. On average, the reaction energy dissipates a 12\% more in high-binding sites. Likewise, $\overline{T}_{\text{COM}}$, N$_{\text{diff}}$, and $\overline{d}$ are significantly higher for MB sites than HB sites. This indicates the importance of different factors in energy dissipation and related properties. 

It is essential to mention that, even though at the end of the dynamics, the reaction energy is not entirely dissipated (see Figure \ref{fig:kinetic_ice}), the translational component $T_{\text{COM}}$ is already wholly dissipated. For WB sites, we observe the highest $\overline{T}_{\text{COM}}$, N$_{\text{diff}}$, and $\overline{d}$ which reinforces the idea that non-thermal events are maximized on WB sites. It is worth mentioning that any trajectory starting from WB, PH migrates to a deep binding site. Moreover, $\overline{\mathcal{F}}$ for WB does not follow the trend that we saw for MB and HB, with  $\overline{\mathcal{F}}$ values in between HB and MB. As we mentioned in Section \ref{sec:exploration}, WB sites are not populated after the reaction, and the reaction energy dissipation takes place either in MB and HB sites following rapid diffusion after formation, and this observation is further reinforced by the values of $\overline{\mathcal{F}}$ starting from WB sites, appearing in between MB and HB values. For $\overline{T}_{\text{COM}}$, we keep finding values corresponding to 1--5\% of the total reaction energy, rarely going as high as 6.8\% in just a single case, in a case starting from a WB situation.

The efficiency of energy dissipation also depends on the hydrogen isotope under consideration. We found that, on average, the energy dissipates 6--7 \% faster when the incoming particle is deuterium, a fact that could be attributed to the coupling of PD with the bending vibration modes of water, as mentioned above. Similarly, we observed that the average diffusion length of the nascent deuterated molecule is lower than for hydrogen additions, by $\sim$0.5 \AA{} on average (which corresponds to a 10\% of the total traversed distance). At this point, it is important to stress that our classical MD simulations disregard the differences that arise from the quantum nature of H and D. For example, ZPVE is not included in our simulations. We do not expect significant changes in the dynamics at short timescales because of how energetic the reaction is. However, quantum effects may be significant during thermalization, and are vital in supporting quantitative claims in the deuterations.

\section{Astrochemical Implications} \label{sec:Discussion}

\subsection{Non-thermal effects of relevance to interstellar chemistry: Astrochemical models}

The findings presented in this paper carry significant implications for the chemistry of phosphorous-bearing species in particular and interstellar chemistry in general. As we presented in the introduction, there is no single detection of phosphorous hydrides, including phosphine (\ce{PH3}), in cold astronomical environments. One common explanation for the absence of a molecule (or family of molecules) in the cold ISM relies on chemical conversions and the appearance of proxy species, a hypothesis applicable to the gas and solid phase \citep{shingledecker_case_2019, Shingledecker2020_b, Cooke2020, Shingledecker2022}). The PH molecule is the most simple phosphorous hydride, and it is thought to be the current onset in the formation of interstellar \ce{PH3} \citep{Chantzos2020, Molpeceres2021b}, but an early release to the gas-phase or efficient non-thermal diffusion to encounter other radicals and react may decrease its abundance. 

Our simulations reveal that between 1 and 5\% of the reaction energy is transferred to translation, a quantity labelled $\chi$ in the literature \citep{Fredon2021}. This number is in agreement with the current formulation of chemical desorption in terms of RRKM theories \citep{Garrod2007, Minissale2016b}, where a value between 0.5 and 1.0 \% is assumed for the fraction of energy going to the desorption mode (e.g., oscillation of the $z$ coordinate of the molecule's centre of mass). Since we are considering three translational degrees of freedom instead of the single one used in the statistical study of chemical desorption, our values are in the range predicted by those models. Our results also fall within the lower bounds considered in the chemical models of \cite{Fredon2021} and specifically with their low-conversion energy model of assumed 5\% of energy conversion. For the specific \ce{P + H -> PH} reaction, 1--5 \% of energy conversion corresponds to 3.14--15.7 kJ mol$^{-1}$ energy available for diffusion or desorption, assuming a univocal value of the reaction energy corresponding to -314 kJ mol${-1}$ (the value in the gas phase).

Regarding non-thermal diffusion, the 1--5 \% of the reaction energy corresponds to more or less the average binding energy of the PH molecule and assuming a diffusion energy is a fraction of that energy (even though such an assumption must be done with care under astrophysical conditions, see \cite{Furuya2022}), the PH molecule can visit several binding sites, reacting in the process with other pre-adsorbed molecules. However, the translation energy may not be enough to overcome any diffusion barriers for reactions starting in high-energy binding sites. This is illustrated by the value of $\overline{d}$, \Tabref{tab:bigSamplingH}. From an astrochemical perspective, one important factor is missing in our simulations: the \ce{H2} molecule coverage on the surface. At 10 K, most high-energy binding sites will be populated by \ce{H2} molecules (10 orders of magnitude more abundant than the most abundant phosphorous bearing molecule \citep{Chantzos2020}), and PH would likely not be locked on high binding sites. Our average diffusing distance ($\overline{d}$) is, therefore, a lower bound of the real one. Hence PH can perform a few hops on the grain, meeting reaction partners and converting to more complex P-bearing molecules. This contrasts with the classical picture of PH$_{x}$ forming under thermal conditions by hydrogenation in which only hydrogen can move. Thus, we suggest that the effective binding energy applicable to non-thermal events should be lower than the average binding energy computed from distributions, which we tested later in our chemical models (see below).

The other critical non-thermal process is chemical desorption, a phenomenon vital for explaining the return of interstellar adsorbates to the gas phase, where they are ultimately detected. Chemical desorption has been treated analytically in the astrochemical literature. Corresponding approaches are briefly reviewed here. The most straightforward scheme assumes a constant desorption probability of normally 1\% for every reaction. The second \cite{Garrod2007} propose a scheme based on RRKM theory explicitly introducing the binding energy and reaction enthalpy in the formalism. Improving on that, \cite{Minissale2016b} also included a collisional parameter ($\epsilon$) where a mass term is introduced, allowing for good agreement for rigid surfaces (graphite) under the assumption of a purely elastic collision. The last attempt to analytically formulate this phenomenon can be found in \cite{Fredon2021}, where the authors explicitly consider the fraction of energy inoculated into translational degrees of freedom ($\chi$), which is a step beyond the equipartition of the reaction energy predicted by RRKM theory. In the latter study, the authors indicate that the least known part of their model is the determination of $\chi$, which is most likely not constant for every reaction. Here, we have found $\chi$ to be between 1 and 5 \% with average values of around 2\% for the \ce{P + H -> PH} reaction (Table \ref{tab:bigSamplingH}), depending on the binding site under consideration. The probability of chemical desorption ($p$) as a function of $\chi$, $\Delta H_\text{r}$ (the enthalpy of the reaction), and EB (the binding energy) proposed in \citet{Fredon2021} for one-product reactions is:

\begin{equation} \label{eq:fredon}
    p = 0.5 \left( 1 - \exp\left(-\dfrac{\chi\Delta H_\text{r} - \text{EB}}{3\text{EB}}  \right) \right).
\end{equation}

\begin{figure}[h!]
    \centering
    \includegraphics[width=8cm]{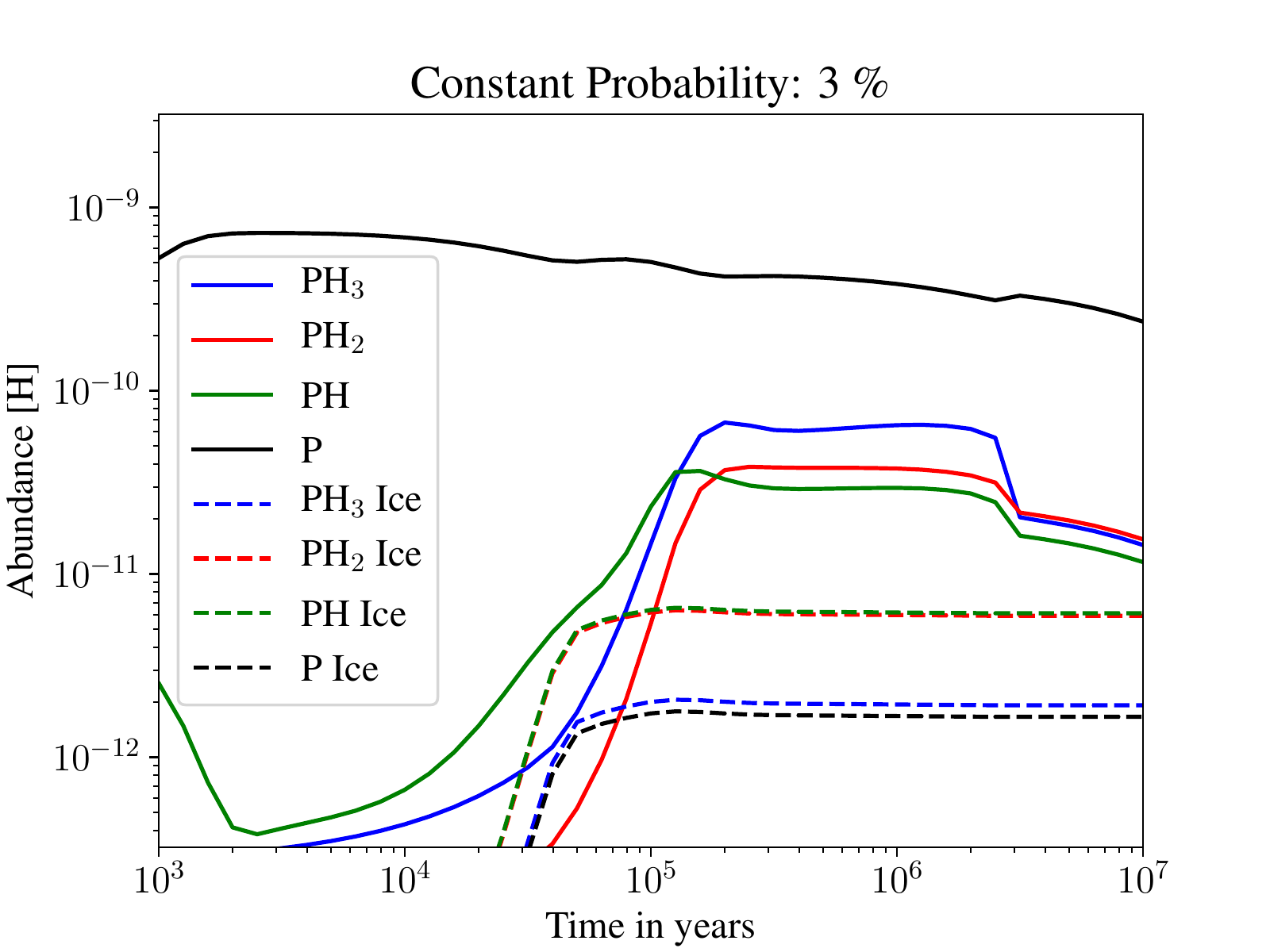}
    \vfill
    \includegraphics[width=8cm]{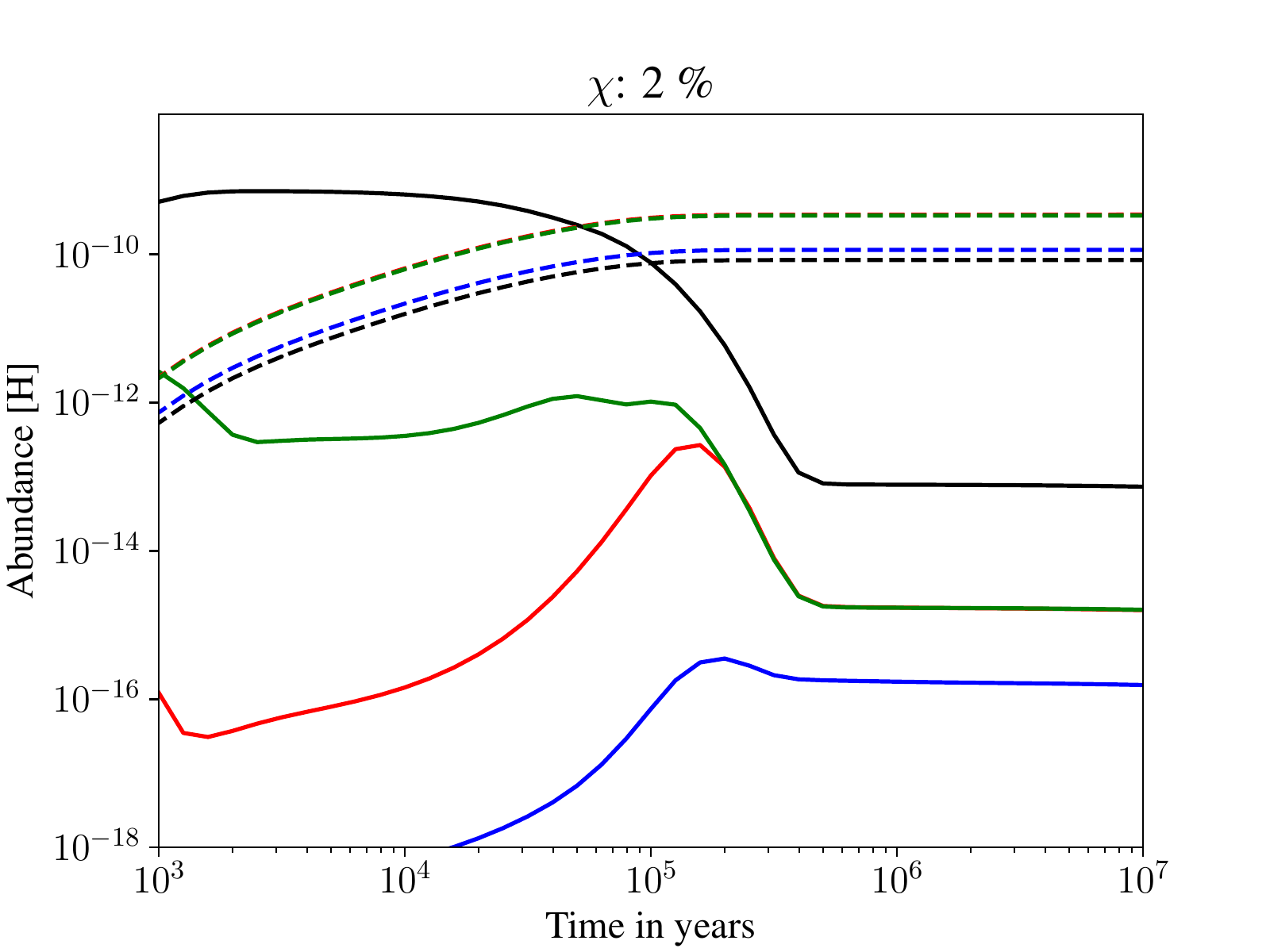}
    \vfill
    \includegraphics[width=8cm]{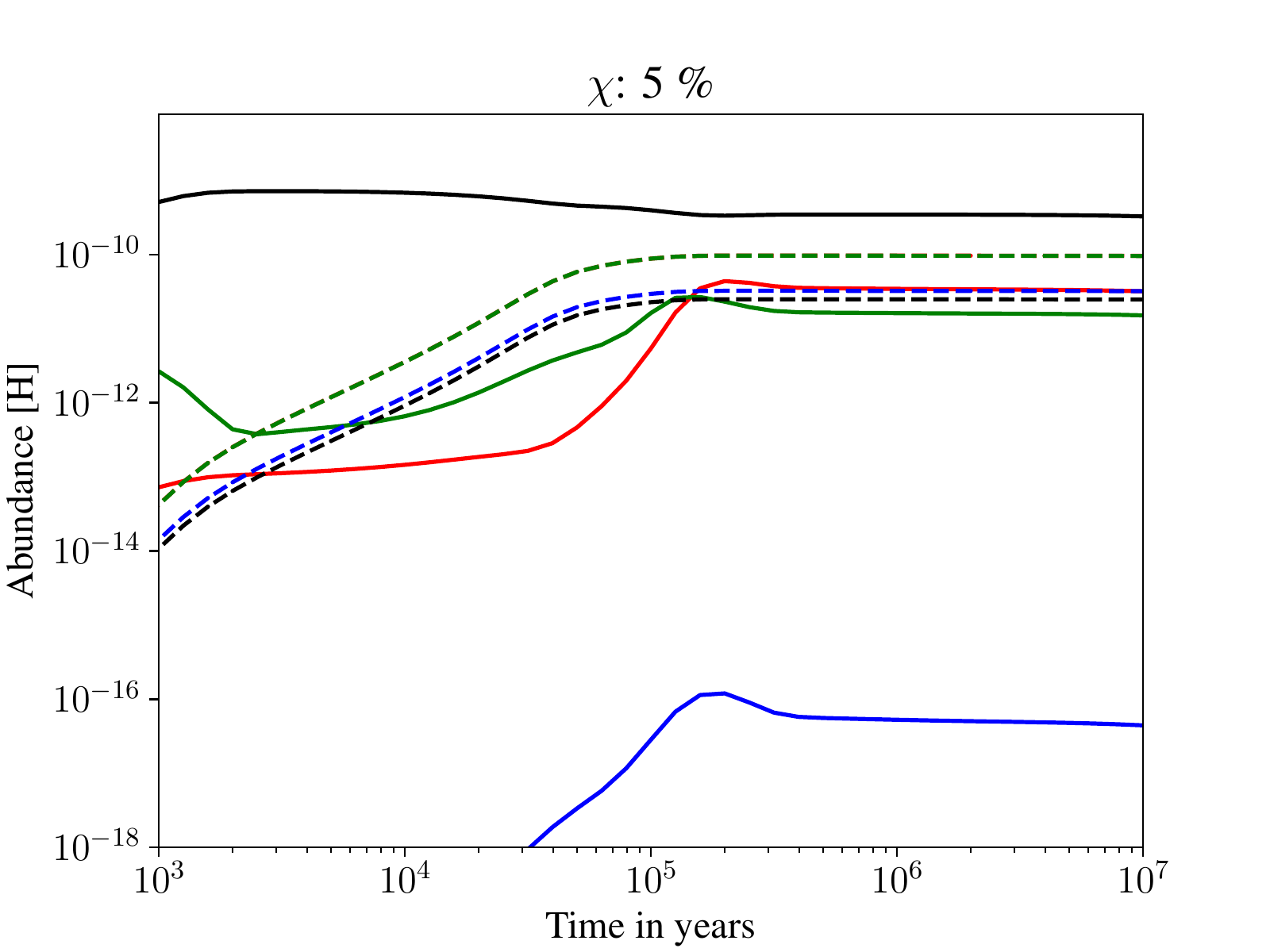}
    \caption{Astrochemical models for the evolution of PH$_{x}$ species in the gas (solid lines) and in the ice (dashed lines) for models with different treatment of chemical desorption. Labels in the topmost graph apply to all graphs.}
    \label{fig:CM1}
\end{figure}

We have explicitly addressed the impact of our derived $\chi$ values in a 3-phase, e.g. gas, grain and (inert) mantle astrochemical model of a molecular cloud. The model is a recent modification of the model reported by \cite{Furuya2022} using the chemical network of \cite{Garrod2013} and adding the hydrogenation sequence of the P atom based on our recent study \cite{Molpeceres2021b} and also the \ce{H2} abstraction reactions for these species. The physical conditions of the model are gathered in Table \ref{tab:conditions_pseudo} and are the standard conditions of a pseudo-time-dependent molecular cloud model. Reaction-diffusion competition is explicitly considered in our models with diffusion energy assumed to be 0.4 of the binding energy. Quantum tunnelling of atomic H is taken into account in the evaluation of the diffusion-reaction competition \citep{changGasgrainChemistryCold2007}. Non-thermal mechanisms are accounted for, including photodesorption (with a constant photodesorption yield per photon of 1$\times$10$^{-3}$), desorption induced by cosmic rays in its Hasegawa and Herbst formulation \citep{Hasegawa1993} with a cosmic-ray-ionization-rate ($\zeta$) of 1.3$\times$10$^{-17}$~s$^{-1}$ and, of most relevance for this paper, chemical desorption. 

To see the impact on the phosphorous hydrides chemistry, we evaluated two chemical desorption schemes, one assuming a constant desorption probability for all phosphorous hydrogenation reactions of $p=3$\% following the recent experimental and modelling studies \citep{Nguyen2021, furuya_quantifying_2022} (and 1\% for all the other reactions) and another one using the \cite{Fredon2021} approach and explicitly including the $\chi$ values derived here for the \ce{P + H -> PH}. Specifically, we took $\chi$ values of 2.0 and 5.0 \%. The former corresponds to the average value obtained in this work, while the latter represents an upper value of it. For simplicity, we assumed that $\chi$ is equal throughout the whole hydrogenation sequence. The $\chi$ value for all the other reactions in the reaction network is set to 1\%. The binding energies for P and \ce{PHx} molecules are taken from \cite{Molpeceres2021b}. The elemental abundances for the model are taken from \cite{Aikawa1999}; \emph{e.g.} corresponding to the low metallicity values, with initial P fractional abundances of 1$\times$10$^{-9}$ with respect to H nuclei, and where all the elements are in atomic or ionic form but H, that is contained in \ce{H2}. Enthalpies of formation used to compute $\Delta H_\text{r}$ are taken from the original from the original \cite{Ruaud2016} reaction network for \ce{P + H -> PH} and \ce{PH + H -> PH2}, based on the KIDA database data \citep{Wakelam2012}. In contrast, for \ce{PH3}, we took it from the NIST database because it was unavailable in the original reaction network.

\begin{table}
    \caption{Initial physical conditions utilized in the molecular cloud model.}
    \label{tab:conditions_pseudo}
    \centering
    \begin{tabular}{cc}
    \toprule
    Parameter & Value  \\
    \midrule
    n$_{\text{gas}}$  & 2x10$^{4}$ cm$^{-3}$    \\
    A$_{v}$           & 10 mag                   \\
    $\zeta$           & 1.3x10$^{-17}$ s$^{-1}$ \\
    T$_g$             & 10 K                   \\
    T$_d$             & 10 K                   \\
    \bottomrule
    \end{tabular}
\end{table}

\begin{table}
    \caption{Gas-phase maximum fractional (with respect to H nuclei) abundances and fractional abundances at t=1$\times$10$^{7}$ yr of PH$_{x}$ molecules in the models shown in Figure \ref{fig:CM1}. The notation A(-B) is equivalent to A$\times$10$^{-B}$.}
    \label{tab:abundances1}
    \centering
\resizebox{\linewidth}{!}{    \begin{tabular}{ccccccc}
    \toprule
     & \multicolumn{3}{c}{Peak Abundances} & \multicolumn{3}{c}
    {t=1$\times$10$^{7}$ yr}\\
  Species & p=3\% & $\chi$=2\% & $\chi$=5\% & p=3\% & $\chi$=2\% & $\chi$=5\% \\
    \midrule
    P & 7.3(-10) & 7.1(-10) & 7.2(-10) & 2.4(-10) & 7.3(-14) & 3.3(-10) \\
    PH & 3.7(-11) & 4.1(-12) & 2.7(-11) & 1.2(-11) & 1.6(-15) & 1.5(-11) \\
    \ce{PH2} & 3.9(-11) & 2.7(-13) & 4.4(-11) & 1.6(-11) & 1.6(-15) & 3.2(-11) \\
    \ce{PH3} & 6.7(-11) & 3.5(-16) & 1.2(-16) & 1.4(-11) & 1.5(-16) & 4.4(-17) \\
    \bottomrule
    \end{tabular}}
\end{table}

The results for the model are presented in Figure \ref{fig:CM1} with peak abundances, and abundances at t=1$\times$10$^{7}$ yr gathered in Table \ref{tab:abundances1}. From the figure, we extract several conclusions: First, we observe that the model with $\chi=2.0$ \% yields gas abundances of P hydrides that are very low, which is partially palliated for $\chi=5.0$ \%. This trend is inverted for ice abundance. In all models explicitly using $\chi$, the abundance of \ce{PH3} is very low, in contrast with the model assuming a constant desorption probability that is supported by recent experiments \citep{Nguyen2021}. We believe that the reduced efficiency for chemical desorption that we found for $\chi$=2\% and 5\% stems from considering a single BE and a single $\chi$ parameter for our systems. From Table \ref{tab:bigSamplingH}, we learned that there is an anti-correlation between BE and $\overline{T}_{\text{COM}}$ (a proxy of $\chi$) with a significant variability for $\chi$. 

\begin{figure}[ht]
    \centering
    \includegraphics[width=8cm]{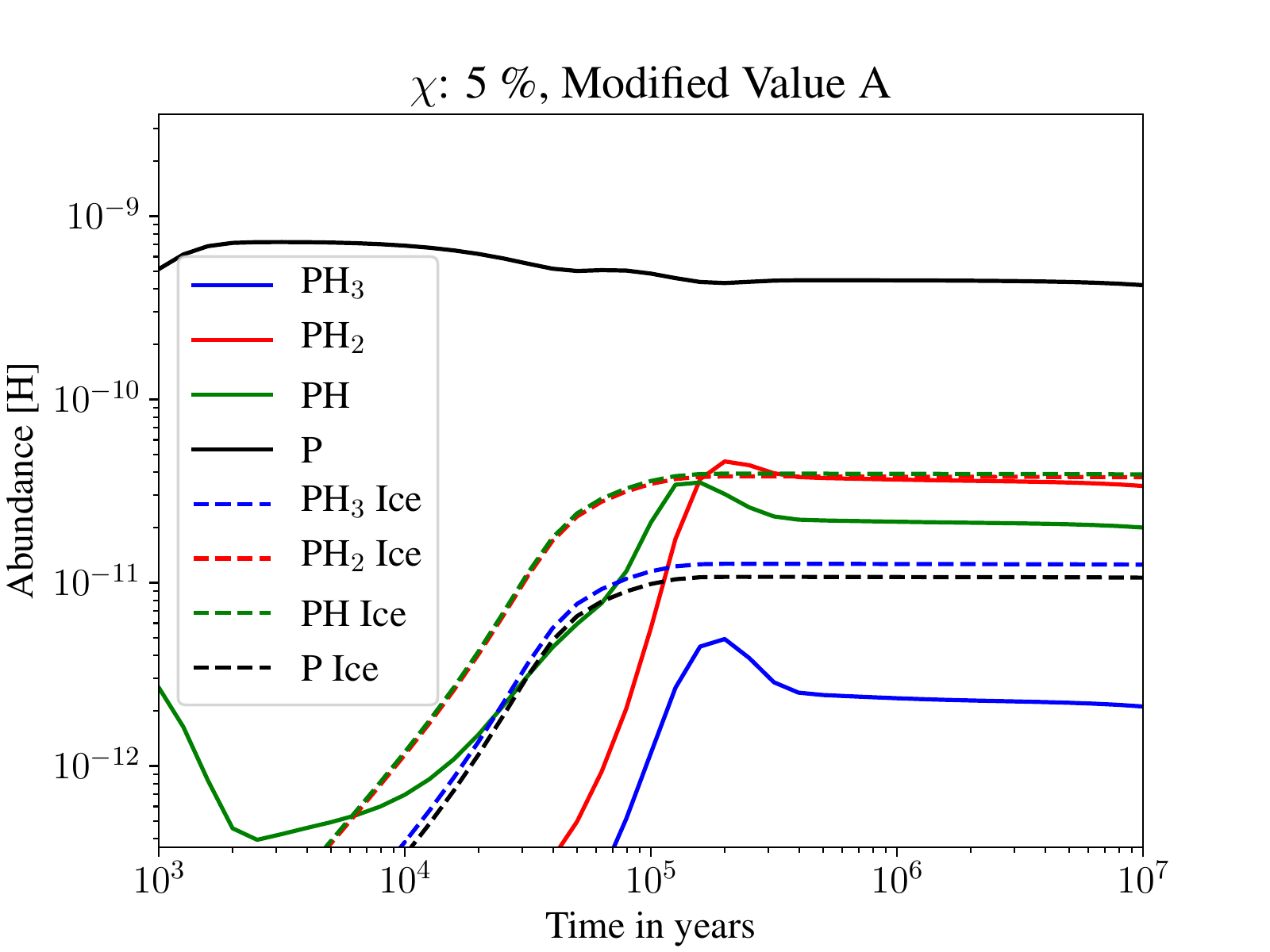}
    \vfill
    \includegraphics[width=8cm]{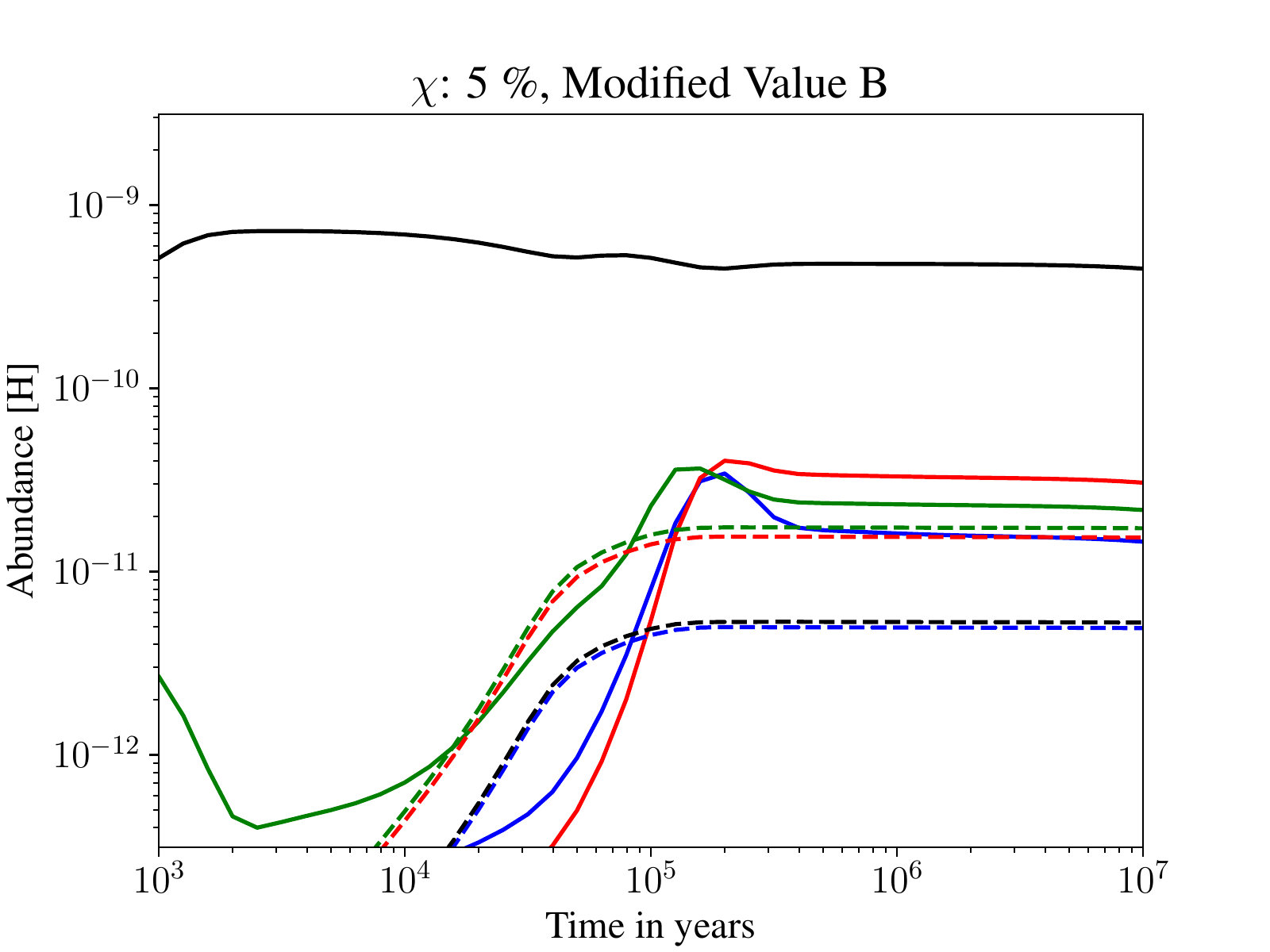}
    \caption{Astrochemical models equivalent to bottom panel in Figure \ref{fig:CM1}, but updating the binding energies according to Table \ref{tab:binding_energies2}. Labels in the topmost graph apply to all graphs.}
    \label{fig:CM2}
\end{figure}

We ran the same chemical models, with $\chi= 5$\% (as an upper value), and using an average BE for PH$_{x}$ that does not includes the high binding sites (Modified Value A), or considering the BE 0.7 of the real binding energy, to consider only weak binding sites (Modified Value B). Hence, the original values of \cite{Molpeceres2021b} are re-weighted according to these rules and included in the model. Table \ref{tab:binding_energies2} enumerates these changes, where for MOdified Value A, we excluded the binding sites with energies higher than 120~\% the average BE. The abundances for this chemical models are shown in Figure \ref{fig:CM2} with peak fractional abundances and abundances at t=1$\times$10$^{7}$ collated in Table \ref{tab:abundances2}.

\begin{table}
    \caption{Update of the PH$_{x}$ binding energies to exclude high-binding energy sites (Modified Value A), and to only consider weak-binding sites (0.70 of EB Modified Value B)}
    \label{tab:binding_energies2}
    \centering
\resizebox{\linewidth}{!}{    \begin{tabular}{cccc}
    \toprule
    Molecule & \cite{Molpeceres2021b} & Modified Value A & Modified Value B  \\
    \midrule
    P            & 1241 & 963  & 869 \\
    PH           & 1616 & 1411 & 1128 \\
    \ce{PH2}     & 1808 & 1523 & 1265 \\
    \ce{PH3}     & 2189 & 1930 & 1532 \\
    \bottomrule
    \end{tabular}}
\end{table}

\begin{table}
    \caption{Same as Table \ref{tab:abundances1} but with the modified binding energies scheme of Table \ref{tab:binding_energies2} and corresponding to the models shown in Figure \ref{fig:CM2}.}
    \label{tab:abundances2}
    \centering
\resizebox{\linewidth}{!}{    \begin{tabular}{ccccc}
    \toprule
     & \multicolumn{2}{c}{Peak Abundances} & \multicolumn{2}{c}
    {t=1$\times$10$^{7}$ yr}\\
  Species &  Model A & Model B & Model A & Model B \\
      \midrule
  P & 7.2(-10) & 7.2(-10) & 4.2(-10) & 4.5(-10) \\
  PH & 3.5(-11) & 3.6(-11) & 1.2(-11) & 2.1(-11) \\
  \ce{PH2} & 4.6(-11) & 4.0(-11) & 3.4(-11) & 3.1(-11) \\
  \ce{PH3} & 4.9(-12) & 3.4(-11) & 2.1(-12) & 1.4(-11) \\
    \bottomrule
    \end{tabular}}
\end{table}

\begin{figure}[ht]
    \centering
    \includegraphics[width=8cm]{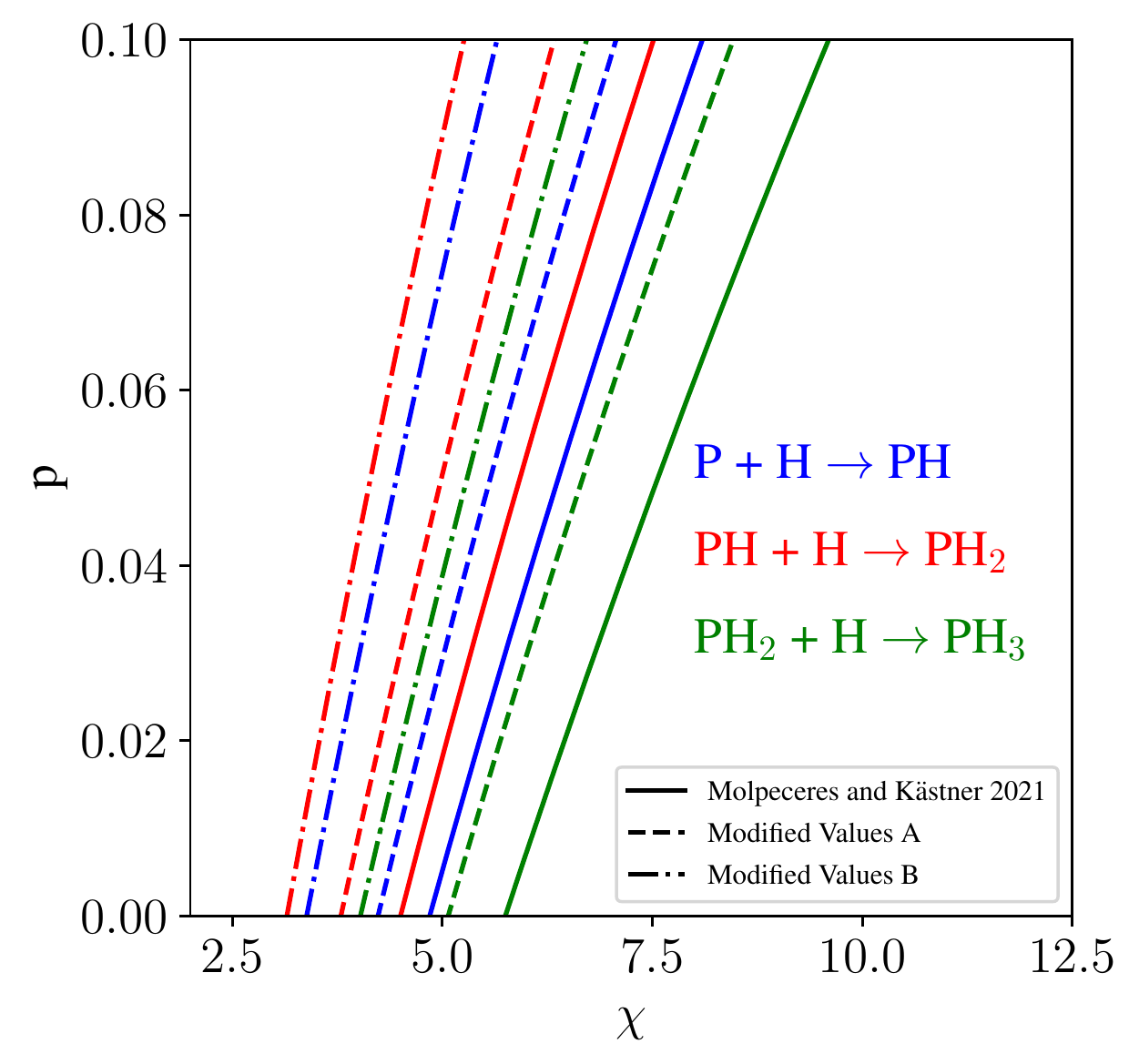}
    \caption{Chemical desorption probability (p) in the \cite{Fredon2021} formulation as a function of the translational energy conversion fraction $\chi$.}
    \label{fig:CM3}
\end{figure}

Even though the changes portrayed in Table \ref{tab:binding_energies2}, Modified Values A, are minimal and fall within the variability for binding energies as presented by various methods for other adsorbates \citep{das_approach_2018, Ferrero2020, bovolenta2022}, we observe a significant increase in the abundance of gas-phase \ce{PH3} by four orders of magnitude. The changes for the Modified Values B are even more drastic because they disregard average and deep binding sites (MB and HB), and we equally observe variations of 5 orders of magnitude in the P-bearing abundances with respect to the values using the original average BE, approaching the values predicted by a constant probability of chemical desorption. This is the key finding of our work because even with an average BE lower than reaction energies by two orders of magnitude, the binding site is still defining the dynamics of chemical desorption of the nascent molecule. For all of our portrayed molecules, both in Figure \ref{fig:CM1} and \ref{fig:CM2}, we can calculate the analytic chemical desorption probability ($p$) as a function of $\chi$. This is shown in Figure \ref{fig:CM3}, and is helpful to rationalize our observations. We see that for the model with the \cite{Molpeceres2021b} BE and $\chi$=2\% (Figure \ref{fig:CM1}, middle panel), the chemical desorption probability is null for all reactions. The molecules' release to the gas phase is driven only by cosmic ray-induced desorption and photodesorption, mechanisms that are insufficient to release a significant portion of the phosphorous hydrides contained in the ice. For $\chi$=5\% (Figure \ref{fig:CM1}, bottom panel) and \ce{PH} and \ce{PH2}, there is a real value of the chemical desorption probabilities that is higher for \ce{PH2} owing to its higher $\Delta H_\text{r}$. Reducing the effective binding energy with the Modified Values B of Table \ref{tab:binding_energies2} finally enables desorption of \ce{PH3} for $\chi$=5 \% (Figure \ref{fig:CM2}, bottom panel).  

A model based on translational energy gained by the nascent molecule, such as the one presented in \cite{Fredon2021}, is a significant step forward in understanding chemical desorption. Furthermore, it is grounded on a physically sound basis, and allow to go beyond equipartition into vibration, rotation and translational degrees of freedom, something that we also do in our simulations. However, in this work, we found that such a model is susceptible to the $\chi$ parameter, which, in addition, is not only species-dependent but also binding-site-dependent. In line with our results, to reproduce the experimental abundances for the chemical desorption probability of \ce{PH3}, \cite{furuya_quantifying_2022} derived a $\chi$ value for chemical desorption of 7\% to account for a chemical desorption probability per reactive event of 3\% in the \ce{PH2 + H -> PH3} reaction, the same value obtained in this work (Continue green line in Figure \ref{fig:CM3}). However, based on our simulations on PH, we cannot explain a value of 7\% for $\chi$, where our upper bound for the distribution is 5\%  with the absolute maximum value found for a single trajectory is 6.7\%, occurring only once, and in a system (PH) with fewer degrees of freedom to distribute the reaction energy into the molecule. The difference between 1--5 \% of our simulations and 7\% obtained from the fit of experimental data stems from the range of values in EB, $\Delta H_\text{r}$, and $\chi$, making it challenging to establish a clear and univocal relation between our microscopically derived data with macroscopical models of chemical desorption. We will continue investigating a way of improving on the basis provided by equation \ref{eq:fredon} and encourage further work in this subject.

Finally, from an astrochemical point of view, we cannot confidently determine the abundances of \ce{PH3} in molecular clouds, but it is not our intention with this paper. We refer the reader to \cite{Chantzos2020} for a much more in-depth analysis of interstellar phosphorous chemistry. The purpose of the here presented chemical models is to evaluate the sensitivity of the models to the parameters governing chemical desorption, reiterating the extreme variability induced by the BE distributions.

\subsection{Deuteration reactions and PD dynamics}

Finally, we dedicate a few words to our study of the energy dissipation of PD \emph{vs} PH. Although we are aware that the main caveat of our simulations remains the classical nature of the nuclear motion, we found that energy dissipation is, on average, more efficient for PD than for PH. It remains to establish if such an effect is a particularity of PD because of a coupling between the fundamental stretching mode of PD and the water bending modes or if, by contrast, is a consequence of the different mass terms during the dynamics. Suppose this effect is due to the coupling between vibrational frequencies. In that case, we can expect a preferential distillation of molecules whose vibrational modes are compatible with the substrate's composition, which will be important for desorption of molecules as a function of the surface composition. In contrast, if the greater energy dissipation is general to all deuterations, that would imply that molecules on grains would be deuterium enriched compared to their gas counterparts. 

Although the limitation of our method does not allow for strong conclusions in this regard, we encourage further work in this direction because of the crucial implications that our simulations carry for the deuterium fractionation in the ISM. We postpone the respective study to our future work. 

\section{Conclusions}

In this work, we have used a novel methodology based on neural-network interatomic potentials to study the chemical energy dissipation in the surface reaction \ce{P + H -> PH} on ASW. The main finding of our work, that is extremely counterintuitive, is that, even with the reaction energy being two orders of magnitude higher than the binding energy, the initial binding energy of the adsorbate prior to reaction is absolutely crucial, to the point that in our study, only adsorbates on WB sites are susceptible of experience chemical desorption. From our study, we draw several conclusions, which we list below:

\begin{enumerate}
    \item Our MLIP achieves high accuracy, comparable to the reference level of theory, even in reactivity studies, where very abrupt changes in the potential occur in very short time scales.
    \item The binding energies of P and PH on ASW, obtained by neural-network potentials, are, on average, 1317 and 1843 K, values in very good agreement with our previous study (1241 and 1616 K).
    \item Our exploratory simulations of four types of binding sites show that below-average binding energy sites are not likely to be populated after reaction on a pristine surface. By contrast, medium and high energy binding sites are likely to be the endpoint of a nascent molecule, where the reaction energy is dissipated.
    \item Non-thermal effects such as diffusion and desorption are determined early in the dynamics, in the first $\sim$ 10 ps. Energy dissipation and non-thermal events seem to be decoupled processes with a marked dependence on the binding energy of the adsorbate.
    \item From a more extensive pool of simulations on high, medium and weak binding energy sites, we could constrain the fraction of dissipated energy and, most importantly, the translational kinetic energy acquired by the PH molecule after the reaction. We showed that on high-energy binding sites, energy dissipates 12\% more than on average binding sites. Concerning transfer to translational energies, we found that it is a rather inefficient process, and we constrained it to a yield of 1--5 \%. Molecules formed on medium-energy binding sites acquire, on average, nearly double the translational energy as molecules formed on high-energy binding sites. In weak binding sites, the energy dissipation fraction is meaningless because the population of weak binding sites is null after the reaction, with a consequential maximum of translational energy. 
    \item We found the differences between the deuteration reaction and the hydrogenation, \ce{P + D -> PD}. On average, energy dissipates 6-7 \% more for the same binding site for the deuteration reaction, and PD acquires around 10\% less translational energy.
    \item We incorporated our derived values into astrochemical models accounting for chemical desorption. We found that the latest literature model for chemical desorption (eq. \ref{eq:fredon}) is extremely sensitive to the molecule's binding energy and the fraction of energy redistributed to translational degrees of freedom. Using our average value, $\chi$=2\%, we obtained that PH$_{x}$ gas abundances must be very low, which confronts recent studies \citep{Nguyen2020, furuya_quantifying_2022}. Using an upper limit of $\chi$ along with updating binding energies to exclude high-energy binding sites (a situation expected in the ISM for P-bearing molecules), we found changes in between 1-5 orders of magnitude in the abundances of PH$_{x}$ compounds, reconciling our values with those of recent studies mentioned above. This strongly suggests that chemical desorption, like thermal diffusion or desorption, is sensitive to the binding site under consideration.
\end{enumerate}

\noindent Continuations to this work will deepen the connection between the atomistic simulations and the analytical formulas for chemical desorption (and diffusion), ideally tackling the problem of using a single binding energy/translational energy value in contrast to the distribution of such values. 

\begin{acknowledgements}
G.M. thanks the Japan Society for the Promotion of Science (Grant P22013) for its support. V.Z. and J.K. acknowledge the support by the Stuttgart Center for Simulation Science (SimTech). Funded by Deutsche Forschungsgemeinschaft (DFG, German Research Foundation) under Germany's Excellence Strategy - EXC 2075 - 390740016. The authors acknowledge support by the state of Baden-Württemberg through bwHPC and the German Research Foundation (DFG) through grant no INST 40/575-1 FUGG (JUSTUS 2 cluster). KF acknowledges support from JSPS KAKENHI grant Numbers 20H05847 and 21K13967. Y.A. acknowledges support by Grant-in-Aid for Scientific Research (S) 18H05222, and Grant-in-Aid for Transformative Research Areas (A) grant Nos. 20H05847.
\end{acknowledgements}

\bibliographystyle{aa}
\bibliography{sample.bib}

\begin{appendix}
\section{Quality assessment for employed MLIPs} \label{sec:appendix}

\begin{figure}
    \centering
    \includegraphics[width=8cm]{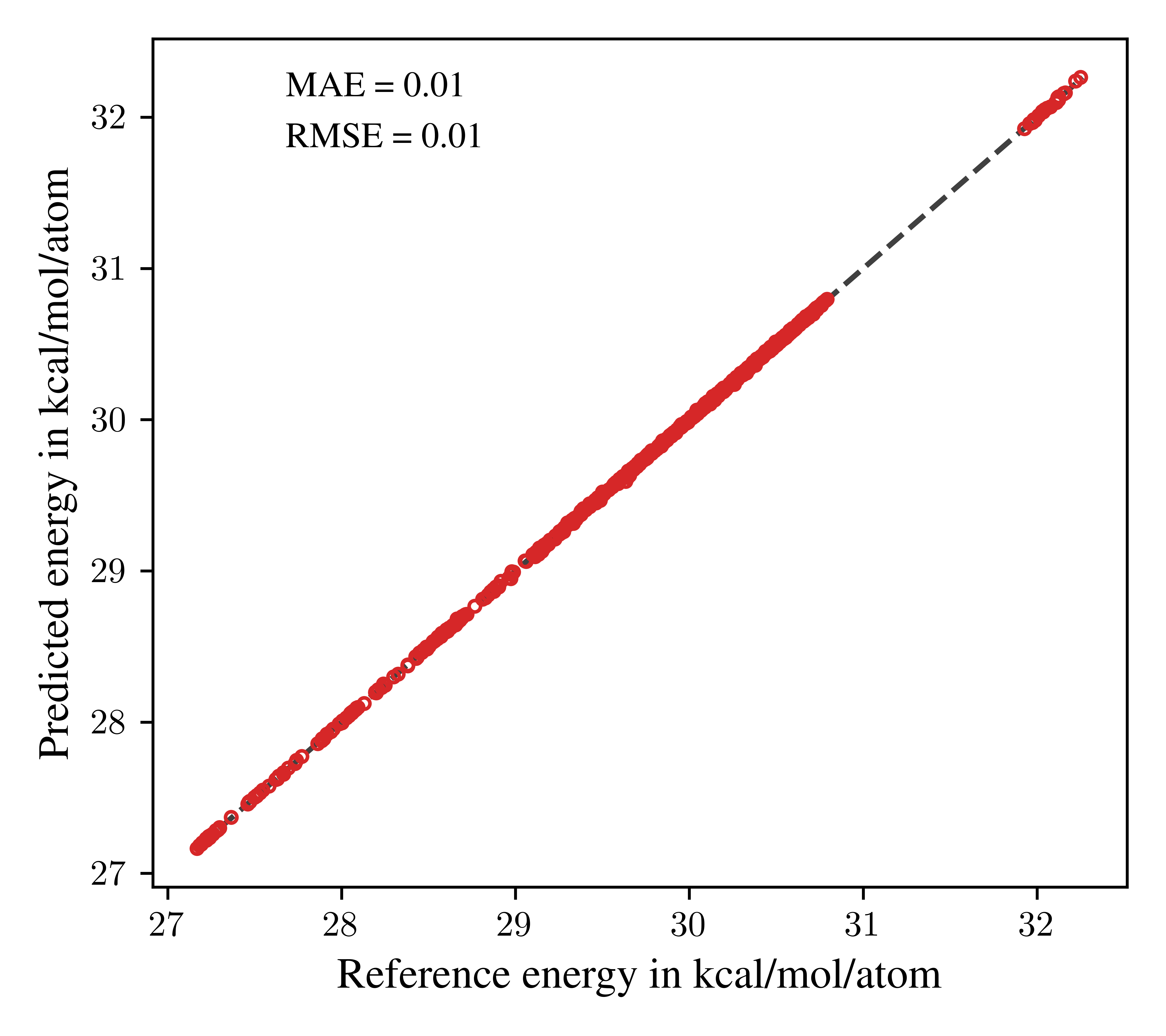}\hfill
    \includegraphics[width=8cm]{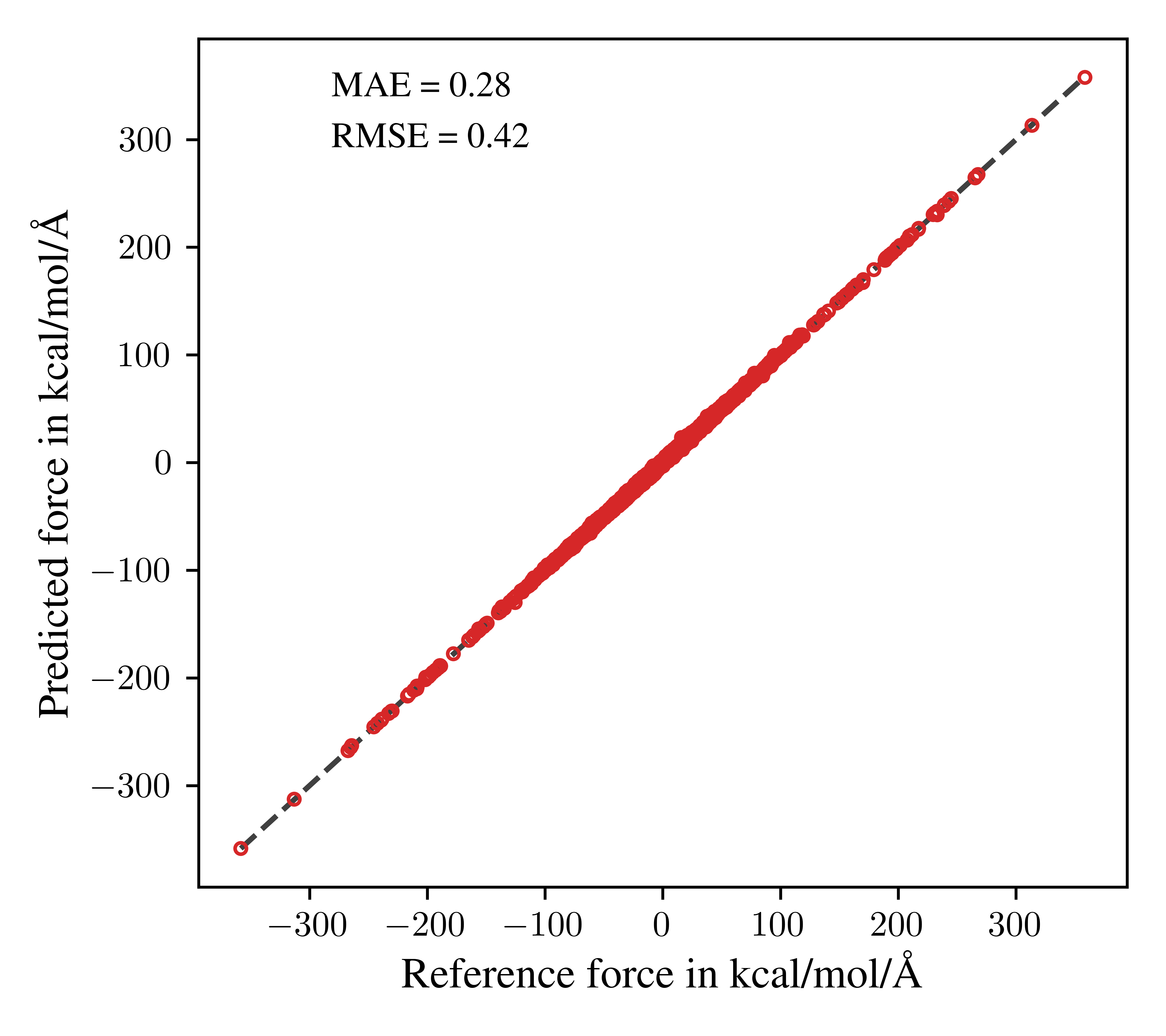}
    \caption{Correlation of the predicted potential energies (Top) and atomic forces (Bottom) with the corresponding reference values for all structures in the test data. Mean absolute (MAE) and root-mean-square (RMSE) errors are shown as an inset.}
    \label{fig:test_errors}
\end{figure}

In this section, we assess the quality of the GMNN-based MLIPs employed to study the reaction dynamics of the \ce{PH} formation. For this purpose, first, we analyse the errors obtained on the test data, i.e., the data drawn from the original data set but not seen during training, including the early stopping technique. \Figref{fig:test_errors} shows the correlation of predicted energies and atomic forces with the respective reference values. All values lie tightly on the diagonal, indicating an excellent performance of developed MLIPs. We obtained a mean absolute error (MAE) of $0.01$\,kcal/mol/atom and $0.28$\,kcal/mol/\AA{} for predicted energies and atomic forces, respectively. The respective root-mean-square errors (RMSE) are $0.01$\,kcal/mol/atom and $0.42$\,kcal/mol/\AA{}.

\begin{figure}
    \centering
    \includegraphics[width=8cm]{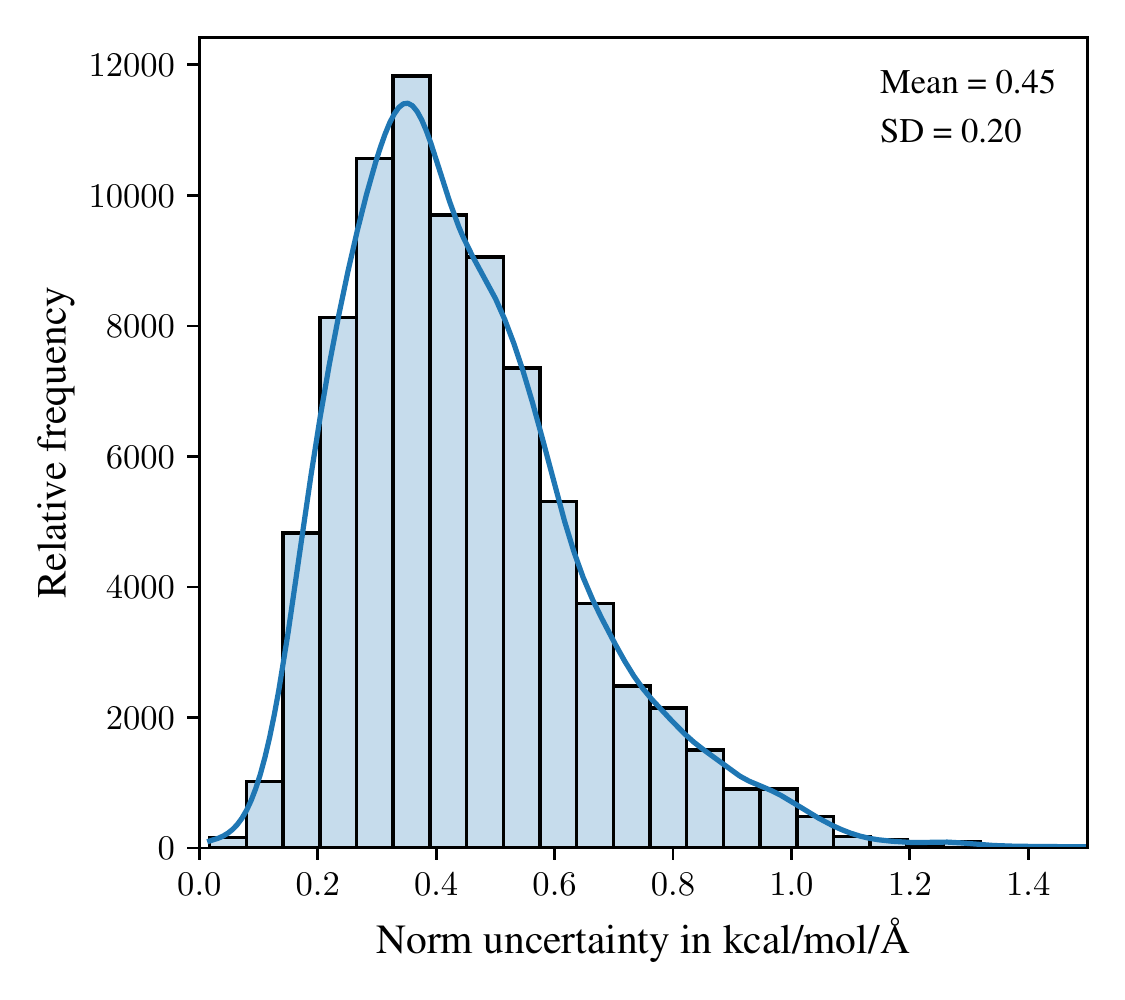}\hfill
    \includegraphics[width=8cm]{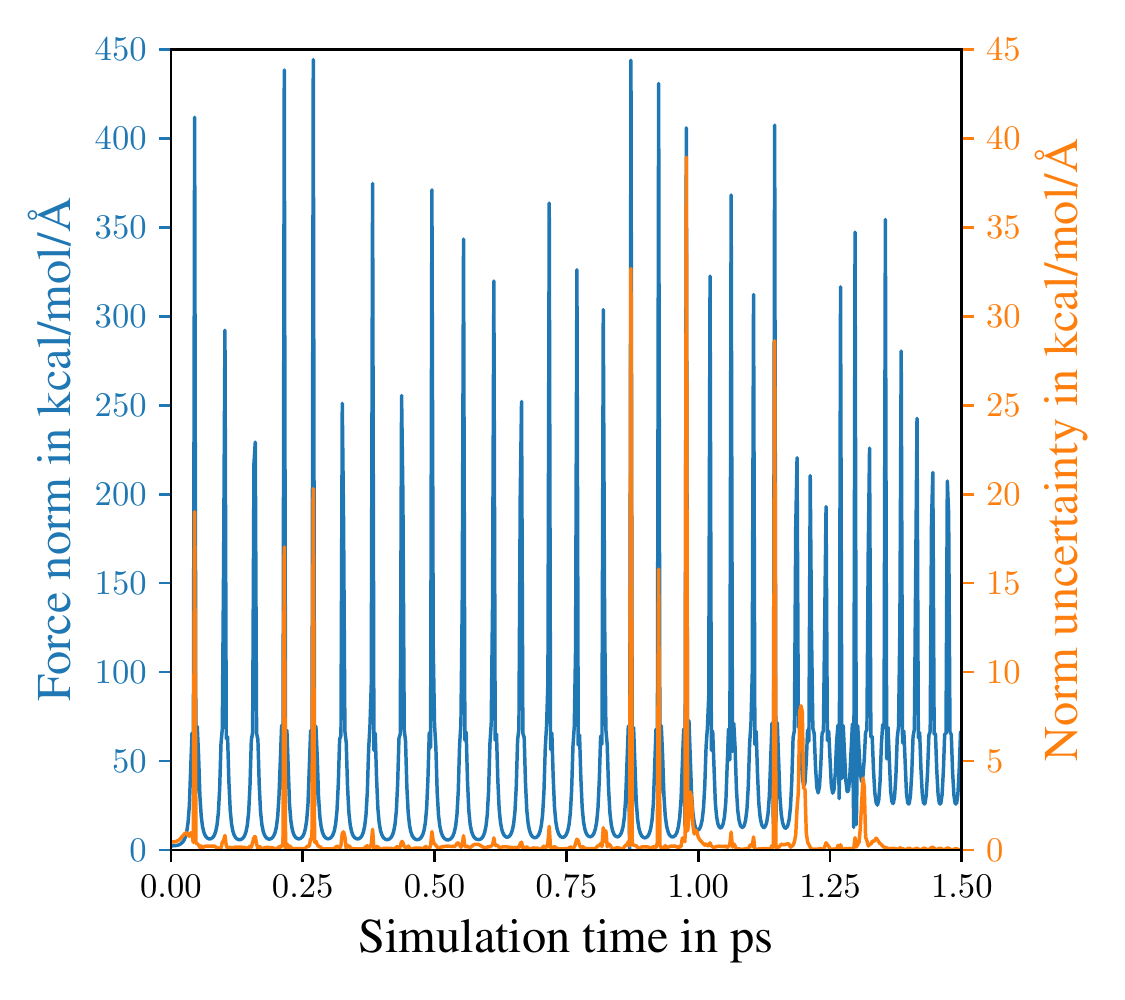}
    \caption{(Top) Norm uncertainty of the atomic forces predicted by an ensemble of three GM-NN models. (Bottom) Comparison of the force norm and the respective uncertainty for the hydrogen atom during first steps of an MD simulation leading to the desorption of the created \ce{PH} species. The time scale is restricted to the reaction of \ce{P} and \ce{H} and the desorption of the resulting \ce{PH} molecule. The uncertainty of the ensemble does not exceed 9.6\,\% of the respective force norm.}
    \label{fig:bounce}
\end{figure}

The MAE and RMSE values in predicted energies and atomic forces are abstract quality measures for MLIPs. To assess the potential's robustness and reliability, we investigate their uncertainty during real-time applications. For this purpose, we consider two possible outcomes, namely a reaction with subsequent desorption \textbf{(1)} and diffusion \textbf{(2)} of created \ce{PH} species, and the distributions of uncertainties corresponding to them. \Figref{fig:bounce} (left) shows the distribution of uncertainties for the case where \ce{PH} desorbs after the reaction \textbf{(1)}, obtained by disagreement between an ensemble of three models. The figure shows that most uncertainties are of values less than 1\,kcal/mol/\AA{}, an empirical threshold used to make statements on the model's reliability. Thus, the model performs overall robustly and leads to correct dynamics. 

\begin{figure}
    \centering
    \includegraphics[width=8cm]{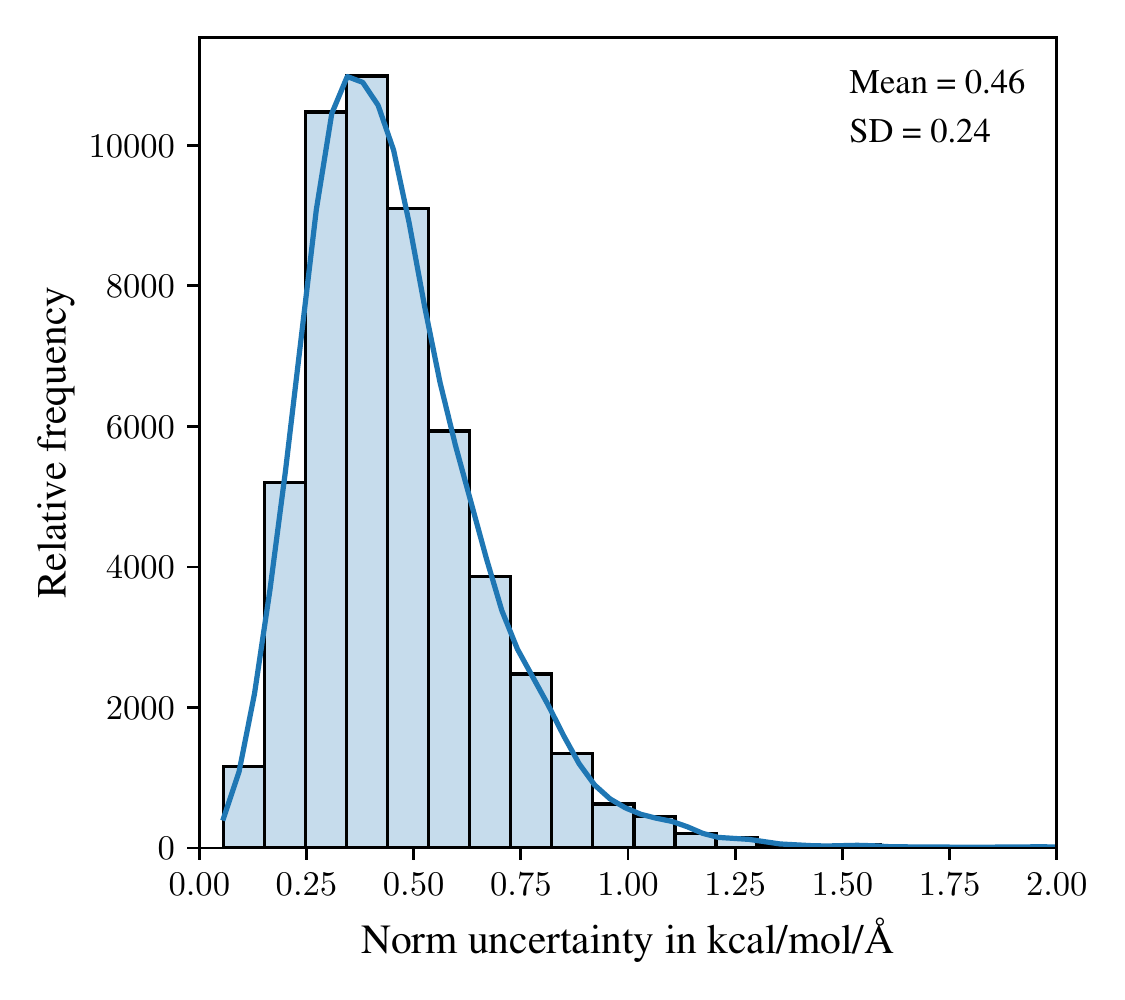}\hfill
    \includegraphics[width=8cm]{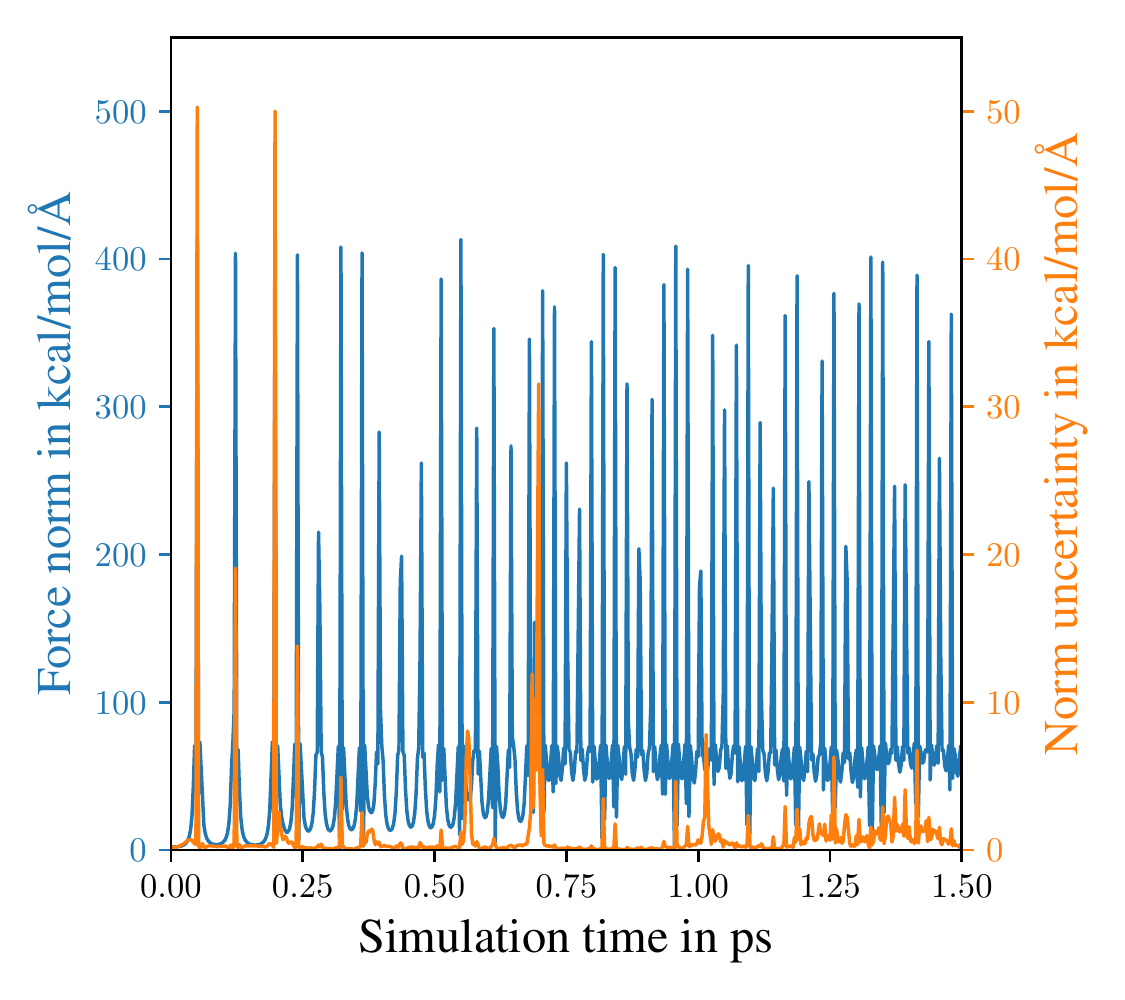}
    \caption{(Top) Norm uncertainty of the atomic forces predicted by an ensemble of three GM-NN models. (Bottom) Comparison of the force norm and the respective uncertainty for the hydrogen atom during first steps of an MD simulation leading to the diffusion of the created \ce{PH} species. The time scale is restricted to the reaction of \ce{P} and \ce{H} and the diffusion of the resulting \ce{PH} molecule. The uncertainty of the ensemble does not exceed 12\,\% of the respective force norm.}
    \label{fig:stick}
\end{figure}

However, at least one of the atoms participating in the reaction requires more rigorous analysis. \Figref{fig:bounce} (right) shows the norm of the force acting on the hydrogen atom evaluated along a molecular dynamics trajectory. Additionally, we show the uncertainty of the force norm in \Figref{fig:bounce} (right). The figure shows that the maximal uncertainty does not exceed 9.6\,\% of the value of the respective force norm. However, additional inspection has shown that those uncertainties correspond to high-energy regions where, e.g., the vibrations stimulated by the reaction should be accounted for. In this case, the distance between P and H reaches values smaller than those sampled by the training set, leading to a higher model's uncertainty. As this part of the potential energy surface is irrelevant to the performed study and its results, the uncertainty in those high-energy regions could not influence them. Similar results have been obtained for the case where \ce{PH} species diffuses on the surface after the reaction \textbf{(2)}. The respective uncertainties are shown in \Figref{fig:stick}.

\begin{figure}
    \centering
    \includegraphics[width=8cm]{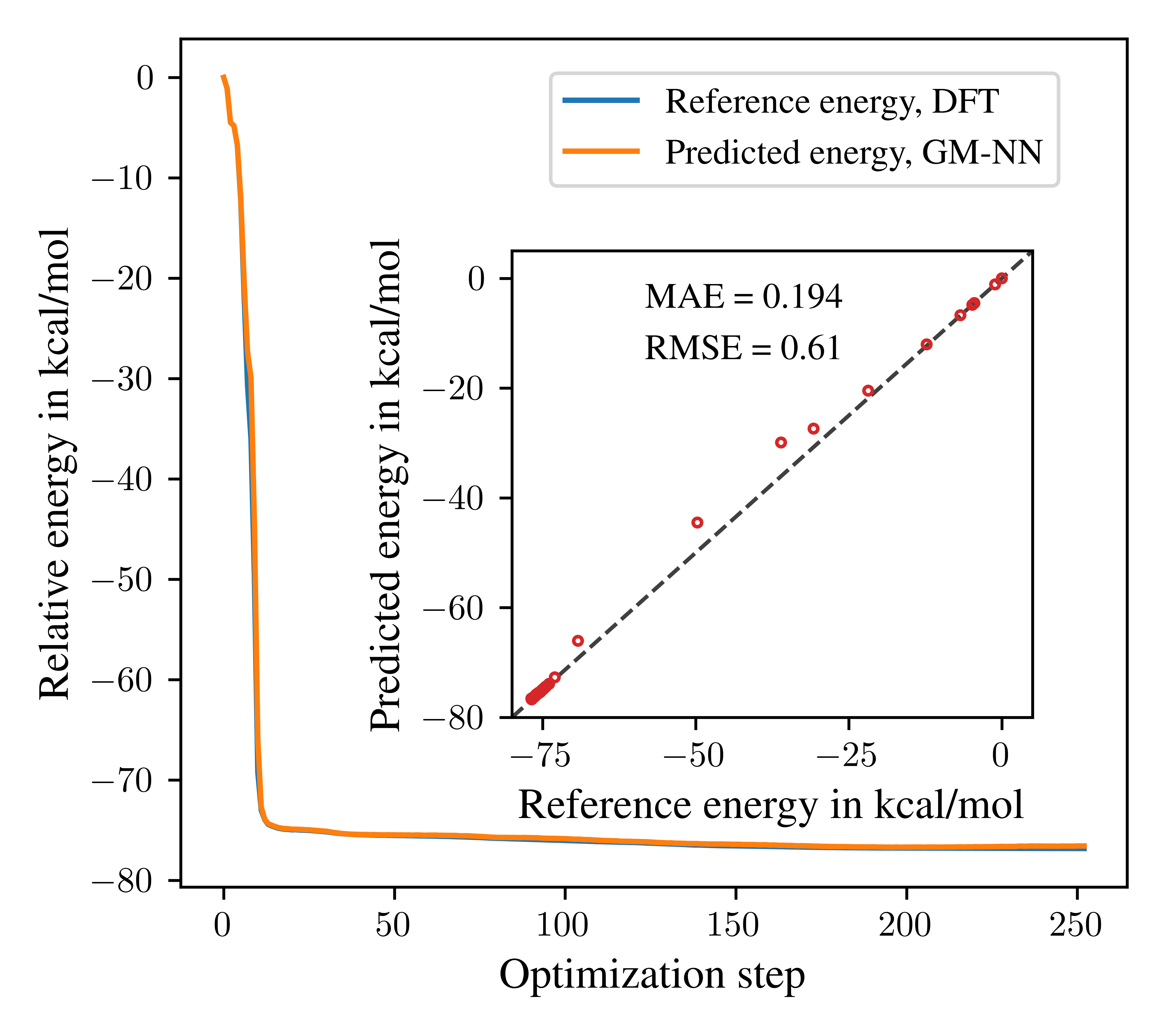}\hfill
    \includegraphics[width=8cm]{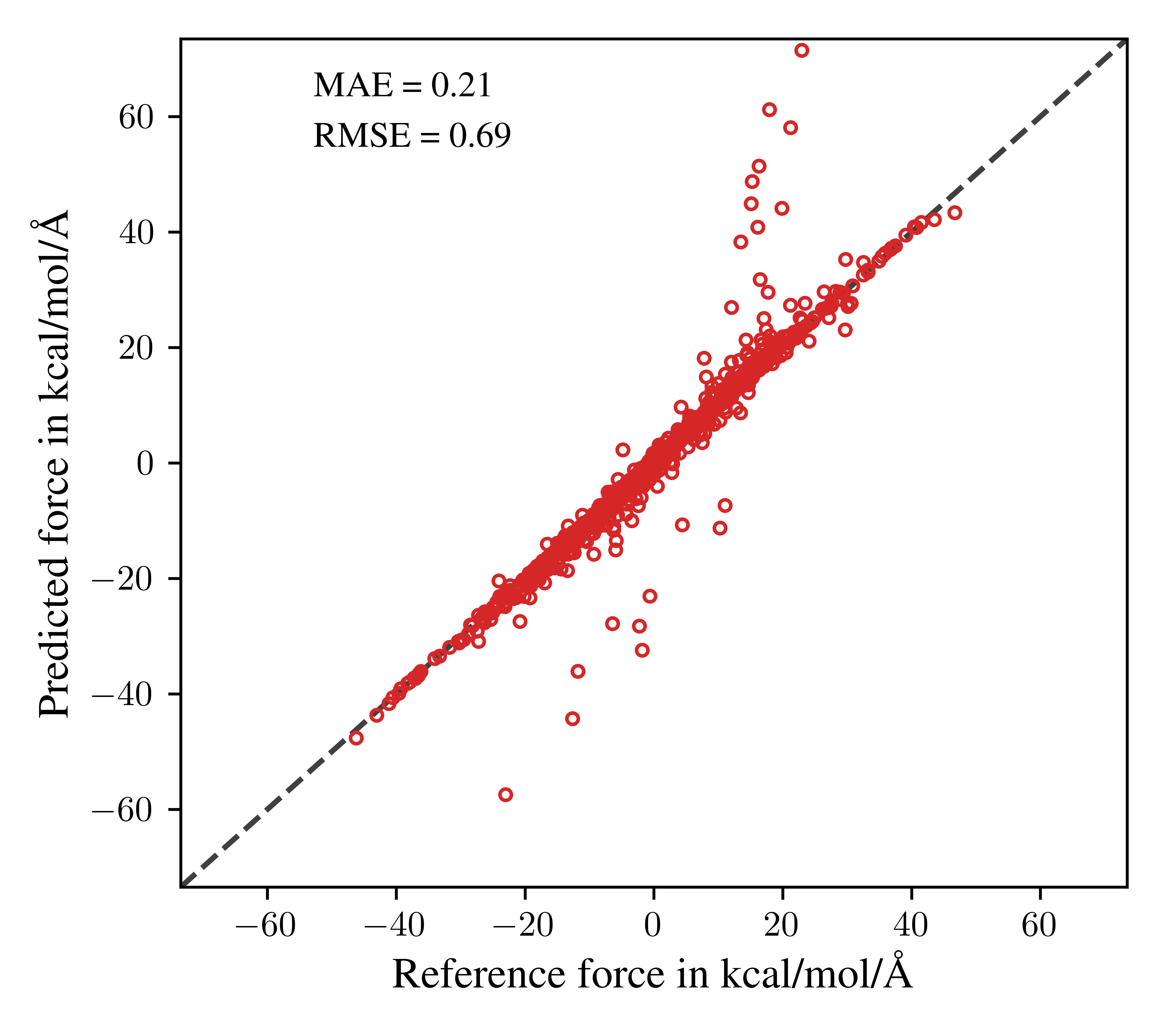}
    \caption{(Top) Comparison of the reference and predicted potential energy profiles obtained during a geometry optimization. (Bottom) Correlation of the atomic forces predicted by the developed MLIP with the corresponding reference values.}
    \label{fig:profile}
\end{figure}

Finally, we consider a geometry optimization profile on a 20 (\ce{H2O}$_{20}$) using explicit DFT calculations at the reference level and using the MLIP. \Figref{fig:profile} shows that both energy profiles almost coincide with an MAE of $0.002$\,kcal/mol/atom and an RMSE of $0.01$\,kcal/mol/atom. Moreover, atomic forces correlate very well with the reference values which corresponds to low MAE and RMSE values of $0.21$\,kcal/mol/\AA{} and $0.69$\,kcal/mol/\AA{}, respectively. This indicates that the MLIP reproduces exactly the reaction profile predicted by DFT.

\end{appendix}

\end{document}